\documentclass[preprint,aps,notitlepage, nofootinbib]{revtex4}
\usepackage{graphicx}
\usepackage{color}
\usepackage{bm}
\usepackage{amsmath,amssymb}
\usepackage{multirow}

\begin{document}
\preprint{OU-HET-946}
\newcommand{\Osaka}{
  Department of Physics, Osaka University,
  Toyonaka 560-0043, Japan
}

\title{
  Atiyah-Patodi-Singer index
  from the domain-wall fermion Dirac operator
}
\author{Hidenori Fukaya}
\author{Tetsuya Onogi}
\author{Satoshi Yamaguchi}
\affiliation{\Osaka}

\begin{abstract}
The Atiyah-Patodi-Singer(APS) index theorem attracts attention for
understanding physics on the surface of materials in topological phases.
The mathematical set-up for this theorem is,  
however, not directly
related to the physical fermion system,
as it imposes on the fermion fields
a non-local boundary condition known as the ``APS boundary condition'' by hand,
which is unlikely to be realized in the materials.
In this work, we attempt to reformulate the APS index in a
``physicist-friendly'' way for a simple set-up
with $U(1)$ or $SU(N)$ gauge group
on a flat four-dimensional Euclidean space.
We find that the same index as APS is obtained from the domain-wall fermion Dirac operator
with a local boundary condition, which is naturally given by the kink structure in the mass term.
As the boundary condition does not depend on the gauge fields,
our new definition of the index is easy to compute with the standard Fujikawa method.
\end{abstract}
\maketitle
\newpage

\section{Introduction}

The Atiyah-Singer(AS) index theorem \cite{Atiyah:1963zz, Atiyah:1968mp}
on a four-dimensional closed
Euclidean manifold $X$ with flat metric is given by
\begin{eqnarray}
  \label{eq:AStheorem}
  n_+-n_- = \frac{1}{32\pi^2} \int_X d^4x\; \epsilon_{\mu\nu\rho\sigma}{\rm tr}_cF^{\mu\nu}F^{\rho\sigma}, 
\end{eqnarray}
where $n_\pm$ denotes the number of $\pm$ chiral zero modes of the Dirac operator $D$,
and $F_{\mu\nu}$ is the field strength of $SU(N)$ or $U(1)$ gauge fields, for which the trace ${\rm tr}_c$ is taken.
This theorem is well known in physics \cite{Jackiw:1977pu}
and can be easily understood
by the so-called Fujikawa method \cite{Fujikawa:1979ay},
\begin{eqnarray}
  n_+-n_- = \lim_{t\to 0}{\rm Tr}\gamma_5 e^{-t D^\dagger D}
  = \lim_{t\to 0} \int d^4 x\; {\rm tr}_{s,c} \sum_n \phi_n^\dagger(x)\gamma_5 e^{-t D^\dagger D} \phi_n(x),
\end{eqnarray}
where the trace ${\rm Tr}$ is taken over space-time coordinates, spinor and color indices, while ${\rm tr}_{s,c}$ means that for spinor and color indices only.
The exponential factor $e^{-t D^\dagger D}$ regularizes the trace (heat kernel regularization).
Taking the simple plane waves for the complete set $\phi_n(x)$,
the right-hand side of Eq.~(\ref{eq:AStheorem}) is obtained as
the leading contribution in the $t$ expansion, 
which survives the $t\to 0$ limit.
Note that the left-hand side is unchanged even when $t$ is finite,
since every non-zero $D^\dagger D$ eigenmode makes a pair with its opposite chirality,
and does not contribute to the trace.

Next let us consider a manifold extending only in the region $x_4>0$,
whose boundary  at $x_4=0$ forms a flat three-dimensional manifold $Y$.
Atiyah, Patodi and Singer (APS) \cite{Atiyah:1975jf} (see also \cite{Atiyah:1976jg, Atiyah:1980jh})
showed that imposing a non-trivial boundary condition (APS boundary condition)
on the Dirac operator, the index is given by
\begin{eqnarray}
  \label{eq:APS}
  \lim_{t\to 0}{\rm Tr}\gamma_5 e^{-t D^\dagger D}
  = \frac{1}{32\pi^2} \int_{x_4>0} d^4x\; \epsilon_{\mu\nu\rho\sigma}{\rm tr}_cF^{\mu\nu}F^{\rho\sigma} -\frac{\eta(iD^{\rm 3D})}{2},
\end{eqnarray}
where $iD^{\rm 3D}$ is the three-dimensional Dirac operator on $Y$,
and $\eta(H)$ is the so-called $\eta$-invariant
which is the (regularized) number of non-negative modes
subtracted by the number of negative modes of a Hermitian operator $H$.
An explicit definition is, for example, given by
the (generalized) $\zeta$-function regularization as
\begin{eqnarray}
\eta(H) = \lim_{s\to 0} \sum_{\lambda \neq 0} \frac{\lambda}{|\lambda|^{1+s}}+h,
\end{eqnarray}
where $\lambda$ denotes the eigenvalue of $H$, and $h$ is the number of zero modes of $H$.
Because of the regularization, $\eta(iD^{\rm 3D})$ is non-integer in general.
In fact, it is equivalent to the Chern-Simons (CS) term  
\begin{eqnarray}
  \frac{\eta(iD^{\rm 3D})}{2} &=& \frac{CS}{2\pi}\;\;\;\mbox{mod integer},\\
  CS &\equiv&  \frac{1}{4\pi}\int_Y d^3 x\; {\rm tr}_c\left[  \epsilon_{\nu\rho\sigma}
    \left(A^\nu \partial^\rho A^\sigma +\frac{2i}{3}A^\nu A^\rho A^\sigma\right)\right],
\end{eqnarray}
which precisely cancels the surface contribution in the first term of Eq.~(\ref{eq:APS}).
Therefore, the total contribution is guaranteed to be an integer.

The APS index theorem describes (a part of) the anomaly descent equations
\cite{Stora:1983ct,Zumino:1983ew,Zumino:1983rz, AlvarezGaume:1983cs, Sumitani:1984ed}. 
The parity anomaly \cite{Redlich:1983dv,Niemi:1983rq,AlvarezGaume:1984nf}
or Chern-Simons term in three dimensions appears as 
the surface term of the axial $U(1)$ anomaly in the bulk four dimensions. 
This (parity) anomaly inflow is important to understand the physics of
topological insulators
\cite{Gromov:2015fda,Witten:2015aba,Metlitski:2015yqa,Seiberg:2016rsg,Tachikawa:2016xvs,
  Freed:2016rqq,Witten:2016cio,Yonekura:2016wuc,Hasebe:2016tjg,Yu:2017uqt}\footnote{
This work is motivated by recent developments in regularization of
chiral fermion using domain-wall fermion formalism,
where the gauge anomaly inflow is manifest
\cite{Grabowska:2015qpk, Fukaya:2016ofi, Okumura:2016dsr, Hamada:2017tny}.
}. 
The APS theorem indicates that
massless edge modes, having parity anomaly,
must appear to cancel the parity violation induced by the $U(1)$ anomaly of bulk fermions.
For this reason, the APS index theorem attracts attention for
understanding physics on the surface of materials in topological phases.

However, the original set-up by APS is not directly related to the physics of topological insulators.
APS considered a Dirac operator for massless fermions with
a non-local boundary condition called the APS boundary condition,
which is introduced in a rather {\it ad hoc} way.
On the other hand, the fermion in a topological insulator is massive in the bulk,
and has a local boundary condition, which keeps
the $SO(3)$ (or $SO(2,1)$ in Minkowski space-time)
rotational symmetry on the surface.
This rotational symmetry is essential for the edge-localized mode
to act as a relativistic Dirac fermion,
but it is not compatible with the helicity conservation,
which is required by the APS condition to keep the bulk fermion massless.
In fact, as we explicitly see below,
the APS boundary condition allows no edge localized mode to exist in the system,
and the eta-invariant appears in an entirely different way
from what we expect in the anomaly inflow between bulk and edge modes.
In this sense, the fact that the APS index describes
the anomaly inflow of topological materials is a coincidence,
since the original mathematical setup by APS is nothing to do with
the physical fermion system.

The goal of this work is to reformulate the APS index
in a ``physicist-friendly'' way, as was done by Fujikawa for the AS index on closed manifolds.
We propose a new index for a fermion Dirac operator with a mass term having a kink structure,
which provides a good model to describe the fermions in topological phases.
We find that this index is identical to the APS index,
which explains why it appears in the anomaly inflow for the topological insulators.
Here, we do not pursue a mathematically precise treatment
but a physically sensible way to do the computation.
For this purpose, we only consider a simple set-up
with the gauge group of $U(1)$ or $SU(N)$ and
flat Euclidean metric both in the four-dimensional
bulk and at the three-dimensional boundaries.
Since the boundary condition does not depend on the gauge fields,
our new definition of the index is easy to compute
with the standard Fujikawa method.

The rest of this paper is organized as follows.
We first review the original APS index theorem in Sec.~\ref{sec:massless},
and discuss the problems of the APS boundary condition when we apply it to physics with boundary.
Then we consider what is required to realize a more physically natural set-up
and show that the domain-wall fermion Dirac operator
\cite{Callan:1984sa, Kaplan:1992bt, Shamir:1993zy, Furman:1994ky} is the best candidate.
We show that the same index as APS is obtained through
the domain-wall fermion Dirac operator in Sec.~\ref{sec:domain-wall} and Sec.~\ref{sec:asymmetricDW}.
Finally we give a summary and discussion in Sec.~\ref{sec:summary}.

\section{Massless fermions with APS boundary condition}
\label{sec:massless}
In this section we reproduce the results by Atiyah {\it et al.} \cite{Atiyah:1975jf}
for a much simpler set-up than the original one.
We consider a massless Dirac operator in the fundamental representation of
$SU(N)$ or $U(1)$ gauge group, taking the $A_4=0$ gauge:
\begin{eqnarray}
  D &=& \gamma_4(\partial_4 + A),
\end{eqnarray}
where $A=\gamma_4 \sum_{i=1}^3 \gamma_i D_i$ with covariant derivative $D_i=\partial_i+iA_i$,
being a Hermitian operator.
We consider a four-dimensional flat manifold $X$ extending in the region $x_4>0$,
with a three-dimensional boundary $Y$ at $x_4=0$.

Then we require the fermion fields,
on which $D$ operates, to have a support only from
negative eigenfunctions of $A$ at the boundary $x_4=0$,
which is known as the APS boundary condition.
Namely, any positive eigenfunction component must vanish:
\begin{eqnarray}
\left.\frac{A+|A|}{2}\phi\right|_{x_4=0} = 0.
\end{eqnarray}
We also consider the opposite case,
\begin{eqnarray}
\left.\frac{A-|A|}{2}\phi\right|_{x_4=0} = 0,
\end{eqnarray}
which we call the anti-APS condition.
Since the spectrum of $A$ requires information of
gauge fields in the entire $Y$, the APS/anti-APS boundary
conditions are non-local.
With these non-trivial boundary  conditions, the anti-Hermiticity of
$D$ is maintained since
\begin{eqnarray}
  (\phi_1, D\phi_2) \equiv \int_X d^4 x \phi_1^\dagger (x) D \phi_2(x)
  = \left.\int_Y d^3 x \phi_1^\dagger (x) \gamma_4 \phi_2(x)\right|_{x_4=0} - (D \phi_1, \phi_2)
  =  - (D \phi_1, \phi_2),
\end{eqnarray}
where we have used the absence of the surface term
\begin{eqnarray}
  \left.\int_Y d^3 x \phi_1^\dagger (x) \gamma_4 \phi_2(x)\right|_{x_4=0} =0,
\end{eqnarray}
which is a consequence of anti-commutation relation $\{\gamma_4, A\}=0$,
so that $\gamma_4 \phi_2$ has a support only from eigenfunction of $A$
with opposite sign of eigenvalues to that of $\phi_1$.
Therefore, their inner product vanishes.

The anti-Hermiticity of the Dirac operator is not enough
to formulate the index theorem since we need twice the operations of $D$
or $D^\dagger D = - D^2$ to regularize the trace of $\gamma_5$.
Therefore, we impose the same APS/anti-APS boundary condition
also on $D\phi$ (then $D^n \phi$ for any $n$ automatically
satisfies the same boundary condition).

In this section, it is convenient to take the chiral representation of the $4\times 4$
gamma matrices or, equivalently, tensor product of $2\times 2$ matrices as
\begin{eqnarray}
  \gamma_{i=1,2,3}&=&\left(\begin{array}{cc}
     & i\sigma_i\\
   -i\sigma_i & 
  \end{array}\right)=-\tau_2\otimes \sigma_i,\;\;\;
  \gamma_4=\left(\begin{array}{cc}
    & 1_{2\times 2}\\
   1_{2\times 2} & 
  \end{array}\right)=\tau_1\otimes 1_{2\times 2},
  \nonumber\\
  \gamma_5&=&-\gamma_1\gamma_2\gamma_3\gamma_4 = \left(\begin{array}{cc}
    1_{2\times 2} & \\
      & -1_{2\times 2}
  \end{array}\right)=\tau_3\otimes 1_{2\times 2},
\end{eqnarray}
where $1_{2\times 2}$ denotes the 2$\times 2$ unit matrix,
and $\sigma_i$ and $\tau_i$ denote the Pauli matrices.
In this representation, $A$ takes a block-diagonal form
\begin{eqnarray}
  A = \left(\begin{array}{cc}
    i D^{\rm 3D} &\\
    & -i D^{\rm 3D}
    \end{array}\right)=\tau_3\otimes iD^{\rm 3D},
\end{eqnarray}
where $D^{\rm 3D}=-\sigma_i D^i$ 
denotes the three-dimensional massless Dirac operator.
Therefore, positive eigenfunctions of $A$
correspond to positive/negative eigenmodes of $iD^{\rm 3D}$ for
positive/negative chiral modes, respectively.
Note that $\gamma_5$ commutes with $A$.
Therefore, these boundary conditions preserve the helicity of the fermions.

There is a crucial difference between the APS and anti-APS boundary conditions.
For simplicity, let us take $A$ as $x_4$ independent.
Then the anti-APS boundary condition allows
an edge-localized zero-mode:
\begin{eqnarray}
        \label{eq:Fermi-Arclikemode}
  \phi =  \phi_\lambda e^{-\lambda x_4},\;\;\; D\phi =0,
\end{eqnarray}
where $\lambda$ and $\phi_\lambda$ are
a positive eigenvalue and eigenfunction of $A$, respectively,
while the  APS boundary condition does not allow such zero modes,
since the sign flip of the eigenvalue $\lambda$ makes the eigenfunction
in Eq.~(\ref{eq:Fermi-Arclikemode}) unnormalizable.

\subsection{Computation on a $x_4$-independent background}

Following the original paper by APS \cite{Atiyah:1975jf}, let us begin with
the case where $A$ and therefore its gauge potentials $A_i$ ($i=1,2,3$) have no $x_4$ dependence.
We take $X$ to be infinitely large in the positive region of $x_4$.
Namely we consider a flat background in the $x_4>0$ region.
In this set-up, $F_{4i}=0$ so that the index theorem
should be simply given as
\begin{eqnarray}
  \label{eq:APScylinder}
  \lim_{t\to 0}{\rm Tr}\gamma_5 e^{-t D^\dagger D}=  -\frac{\eta(iD^{\rm 3D})}{2}.
\end{eqnarray}
The goal of this subsection is to reproduce this result
in our familiar language in physics.

When $A$ has no $x_4$ dependence, $D^\dagger D$ can be written as
\begin{eqnarray}
  D^\dagger D = -\partial_4^2 + A^2,
\end{eqnarray}
which commutes with both $\gamma_5$ and $A$.
It is, therefore, convenient to consider the eigenvalue problem of
$D^\dagger D$ by assuming the form of the solution as
\begin{eqnarray}
  \label{eq:EOM1}
  \phi_\pm (x_4)\otimes\phi_\lambda^{\rm 3D}(\vec{x}),\;\;\; (-\partial_4^2 + \lambda^2)\phi_\pm (x_4)=\Lambda^2 \phi_\pm (x_4),
\end{eqnarray}
where $\phi_\lambda^{\rm 3D}(\vec{x})$ is the eigenfunction of $iD^{\rm 3D}$ with the eigenvalue $\lambda$,
and $\tau_3\phi_\pm (x_4)=\pm \phi_\pm (x_4)$ represent the $\pm$ chiral modes.
The APS boundary condition is expressed by
\begin{eqnarray}
  \phi_+ (x_4)|_{x_4=0}&=&0, \;\;\; (\partial_4-\lambda)
  \phi_- (x_4)|_{x_4=0}=0,\;\;\;\mbox{for $\lambda\geq 0$},\\
  \phi_- (x_4)|_{x_4=0}&=&0, \;\;\; (\partial_4+\lambda)
  \phi_+ (x_4)|_{x_4=0}=0,\;\;\;\mbox{for $\lambda<0$}.
\end{eqnarray}

Let us solve the equation Eq.~(\ref{eq:EOM1}) for the case  $\lambda\geq 0$.
One immediately obtains
\begin{eqnarray}
  \phi^\omega_+ (x_4) &=& \frac{u_+}{\sqrt{2\pi}}\left(e^{i\omega x_4}-e^{-i\omega x_4}\right),\nonumber\\
  \phi^\omega_- (x_4) &=& \frac{u_-}{\sqrt{2\pi(\omega^2+\lambda^2)}}
  \left((i\omega+\lambda)e^{i\omega x_4}+(i\omega-\lambda)e^{-i\omega x_4}\right),
\end{eqnarray}
where
\begin{eqnarray}
  u_+= \left(
  \begin{array}{c}
      1\\
      0
  \end{array}
  \right),\;\;\;
   u_-= \left(
  \begin{array}{c}
      0\\
      1
  \end{array}
  \right),
\end{eqnarray}
and $\omega=\sqrt{\Lambda^2-\lambda^2}$. Both solutions satisfy
\begin{eqnarray}
  \label{eq:innerprod}
  \int_0^\infty dx_4 [\phi^{\omega'}_\pm(x_4)]^\dagger\phi^{\omega}_\pm(x_4) = \delta(\omega-\omega'),\;
\end{eqnarray}
for positive $\omega$ and $\omega'$.
They also satisfy in a subspace where $iD^{\rm 3D}$ takes the eigenvalue $\lambda$,
\begin{eqnarray}
  \label{eq:completeset}
\sum_{g=\pm}\int_0^\infty d\omega [\phi^\omega_g (x_4)][\phi^\omega_g (x_4')]^\dagger 
=\delta(x_4-x_4')1_{2\times 2},\;
\end{eqnarray}
for $x_4,x_4'>0$.
Namely, $\phi^\omega_\pm  (x_4)$ forms a complete set in the $x_4$ direction
for each eigenmode of three-dimensional operator $iD^{\rm 3D}$.
Note that $\Lambda^2>\lambda^2$ is always required so that no edge-localized zero mode is allowed to exist.

Next, let us compute the kernel of the operator $\gamma_5e^{-t D^\dagger D}$
using the complete set $\phi^\omega_\pm  (x_4)$ obtained above for each $\lambda$.
The $++$ component is a simple Gaussian integral leading to
\begin{eqnarray}
  \langle x_4;+| \gamma_5e^{-t D^\dagger D}| x_4';+\rangle &=&
  \int_0^\infty d\omega e^{-t(\omega^2+\lambda^2)}[\phi^\omega_+ (x_4)]_+
  [\phi^\omega_+ (x_4')]_+^*\nonumber\\
  &=& \frac{e^{-\lambda^2 t}}{\sqrt{4\pi t}}\left[e^{-\frac{(x_4-x_4')^2}{4t}}-e^{-\frac{(x_4+x_4')^2}{4t}}\right],
\end{eqnarray}
where we have used the bracket notation $[\phi^\omega_\pm  (x_4)]_g = \langle x_4; g| \omega;\pm \rangle$
(note here that $[u_\pm]_g$ has nonzero component only for $g=\pm$).
The $--$ component needs a little trick to evaluate,
\begin{eqnarray}
        \langle x_4;-| \gamma_5e^{-t D^\dagger D}| x_4';-\rangle &=&
  \int_0^\infty d\omega e^{-t(\omega^2+\lambda^2)}[\phi^\omega_- (x_4)]_- [\phi^\omega_- (x_4')]_-^*\nonumber\\
  &=& \frac{e^{-\lambda^2 t}}{\sqrt{4\pi t}}\left[e^{-\frac{(x_4-x_4')^2}{4t}}+e^{-\frac{(x_4+x_4')^2}{4t}}\right]+I(x_4+x_4'),
\end{eqnarray}
where 
\begin{eqnarray}
  I(x_4+x_4') &=& e^{-t\lambda^2}\int_0^\infty \frac{d\omega}{2\pi} e^{-t\omega^2}
  \left[\frac{-2i\lambda(\omega-i\lambda)}{\omega^2+\lambda^2}e^{i\omega (x_4+x_4')}+h.c.\right]
  \nonumber\\&=& e^{-t\lambda^2}\int_{-\infty}^\infty \frac{d\omega}{2\pi} e^{-t\omega^2}
  \left[\frac{-2i\lambda}{\omega+i\lambda}e^{i\omega (x_4+x_4')}\right].
\end{eqnarray}
Note here that the integrand has a pole at $\omega=i\lambda$.
In fact, this pole is the origin of the $\eta$ invariant.
$I(x_4+x_4')$ satisfies a differential equation
\begin{eqnarray}
\label{eq:Ieq}
  \left(\frac{\partial}{\partial x_4}-\lambda\right)I(x_4+x_4') =
  \frac{\lambda}{\sqrt{\pi t}}e^{-t\lambda^2}e^{-\frac{(x_4+x_4')^2}{4t}}.
\end{eqnarray}
Here, the solution of Eq.~(\ref{eq:Ieq}) is given by
\begin{eqnarray}
  I(x_4+x_4') = -\lambda e^{\lambda(x_4+x_4')}
  \left\{\mbox{erfc}\left(\frac{x_4+x_4'}{2\sqrt{t}}+\lambda \sqrt{t}\right)+c\right\},
\end{eqnarray}
where the function $\mbox{erfc}$ denotes the complementary error function,
\begin{eqnarray}
  \mbox{erfc}(x)= \frac{2}{\sqrt{\pi}}\int_x^\infty d\xi e^{-\xi^2}.
\end{eqnarray}
Since it takes $\mbox{erfc}(-\infty)=2$, $\mbox{erfc}(0)=1$, and $\mbox{erfc}(\infty)=0$,
the constant $c$ must be zero in order to satisfy the $t\to 0$ limit
converging to Eq.~(\ref{eq:completeset}).

For $\lambda<0$, we obtain the same formula but with $\lambda$ and $\phi^\omega_\pm$ being replaced by
$-\lambda$, and $\phi^\omega_\mp$.
Combining these results, the kernel is evaluated as
\begin{eqnarray}
  \sum_{g=\pm}\langle x_4;g| \gamma_5e^{-t D^\dagger D}| x_4';g\rangle
  &=& {\rm sign} \lambda \left[- \frac{e^{-\lambda^2 t}}{\sqrt{\pi t}}e^{-\frac{(x_4+x_4')^2}{4t}}+|\lambda| e^{|\lambda|(x_4+x_4')}
  \mbox{erfc}\left(\frac{x_4+x_4'}{2\sqrt{t}}+|\lambda| \sqrt{t}\right)\right].
 \nonumber\\
\end{eqnarray}
We can compute the index by taking a trace over
$x$ and $\lambda$,
\begin{eqnarray}
  {\rm Tr}\gamma_5e^{-t D^\dagger D}  
  &=& \sum_\lambda {\rm sign} \lambda \int dx_4 \frac{\partial}{\partial x_4}\left[ \frac{1}{2}e^{2|\lambda|x_4}
    \mbox{erfc}\left(\frac{x_4}{\sqrt{t}}+|\lambda| \sqrt{t}\right)\right] \int_Y d^3y |\phi^{\rm 3D}_\lambda (\vec{y})|^2
  \nonumber\\
  &=& -\sum_\lambda \frac{{\rm sign}\lambda}{2} \mbox{erfc}\left(|\lambda| \sqrt{t}\right).
\end{eqnarray}
Taking the $t=0$ limit, we obtain the desired formula,
\begin{eqnarray}
  \label{eq:etaflatAPS}
  \lim_{t\to 0}{\rm Tr}\gamma_5e^{-t D^\dagger D}= -\sum_\lambda \frac{{\rm sign}\lambda}{2} = - \frac{\eta(iD^{\rm 3D})}{2}.
\end{eqnarray}
It is important to note again that the APS boundary condition
allows no edge-localized modes.
The eta-invariant appears from a non-trivial $\omega$ integration over the bulk modes,
which looks very different from what we expect in physics of topological insulators.


\subsection{Anti-APS boundary condition}

It is interesting to consider the anti-APS boundary condition
for the same set-up, where $A$ has no $x_4$ dependence.
As mentioned before, the crucial difference from the APS boundary condition
is the existence of the edge-localized modes.
The condition
\begin{eqnarray}
  \phi_- (x_4)|_{x_4=0}&=&0, \;\;\; (\partial_4+\lambda)
  \phi_+ (x_4)|_{x_4=0}=0,\;\;\;\mbox{for $\lambda\geq 0$},\\
  \phi_+ (x_4)|_{x_4=0}&=&0, \;\;\; (\partial_4-\lambda)
  \phi_- (x_4)|_{x_4=0}=0,\;\;\;\mbox{for $\lambda<0$},
\end{eqnarray}
allows the edge-localized chiral zero modes,
\begin{eqnarray}
  \phi^{\rm edge}_+ (x_4) &=& u_+\sqrt{2\lambda}e^{-\lambda x_4},\;\;\;\mbox{for $\lambda\geq 0$},\\
  \phi^{\rm edge}_- (x_4) &=& u_-\sqrt{2|\lambda|}e^{\lambda x_4},\;\;\;\mbox{for $\lambda< 0$},
\end{eqnarray}
which satisfy $D^\dagger D  \phi^{\rm edge}_\pm (x_4) = 0$.

As in the previous section, let us compute the case $\lambda\geq 0$.
First, we note that the edge-localized zero mode is isolated from
the bulk nonzero modes,
\begin{eqnarray}
  \phi^\omega_+ (x_4) &=& \frac{u_+}{\sqrt{2\pi(\omega^2+\lambda^2)}}
  \left((i\omega-\lambda)e^{i\omega x_4}+(i\omega+\lambda)e^{-i\omega x_4}\right),\nonumber\\
  \phi^\omega_- (x_4) &=& \frac{u_-}{\sqrt{2\pi}}\left(e^{i\omega x_4}-e^{-i\omega x_4}\right),
\end{eqnarray}
where $\omega=\sqrt{\Lambda^2-\lambda^2}$ must be a real number.
In fact, in contrast to the $-$ chirality sector,
the completeness in the $+$ chirality sector is not 
achieved by the bulk nonzero modes $\phi^\omega_+$ alone,
\begin{eqnarray}
  \int_0^\infty d\omega
  [\phi^\omega_+ (x_4)]_+ [\phi^\omega_+ (x_4')]_+^* =  \delta(x_4-x_4')-2\lambda e^{-\lambda (x_4+x_4')},\;
\end{eqnarray}
whose second term is only canceled by adding $[\phi^{\rm edge}_+ (x_4)]_+[\phi^{\rm edge}_+ (x_4')]_+^*$.

Next, let us compute the kernel of the operator $\gamma_5e^{-t D^\dagger D}$,
\begin{eqnarray}
  \sum_{g=\pm}\langle x_4;g| \gamma_5e^{-t D^\dagger D}| x_4';g\rangle
  &=& \int_0^\infty d\omega e^{-t(\omega^2+\lambda^2)}[\phi^\omega_+ (x_4)]_+ [\phi^\omega_+ (x_4')]_+^*
  + [\phi^{\rm edge}_+ (x_4)]_+[\phi^{\rm edge}_+ (x_4')]_+^*
  \nonumber\\&&
  - \int_0^\infty d\omega e^{-t(\omega^2+\lambda^2)}[\phi^\omega_- (x_4)]_- [\phi^\omega_- (x_4')]_-^*.
\end{eqnarray}
The second and third terms are easily obtained,
\begin{eqnarray}
  \int_0^\infty d\omega e^{-t(\omega^2+\lambda^2)}[\phi^\omega_- (x_4)]_- [\phi^\omega_- (x_4')]_-^*
  &=& \frac{e^{-\lambda^2 t}}{\sqrt{4\pi t}}\left[e^{-\frac{(x_4-x_4')^2}{4t}}-e^{-\frac{(x_4+x_4')^2}{4t}}\right],\\
{}  [\phi^{\rm edge}_+ (x_4)]_+ [\phi^{\rm edge}_+ (x_4')]_+^* &=& 2\lambda e^{-\lambda (x_4+x_4')},
\end{eqnarray}
while the first term becomes
\begin{eqnarray}
  \int_0^\infty d\omega e^{-t(\omega^2+\lambda^2)}[\phi^\omega_+ (x_4)]_+[ \phi^\omega_+ (x_4')]_+^*
  = \frac{e^{-\lambda^2 t}}{\sqrt{4\pi t}}\left[e^{-\frac{(x_4-x_4')^2}{4t}}+e^{-\frac{(x_4+x_4')^2}{4t}}\right]+I'(x_4+x_4'),
\end{eqnarray}
where 
\begin{eqnarray}
  I'(x_4+x_4') = -\lambda e^{-\lambda(x_4+x_4')}
  \mbox{erfc}\left(-\frac{x_4+x_4'}{2\sqrt{t}}+\lambda \sqrt{t}\right).
\end{eqnarray}
For $\lambda<0$, we obtain the same formula but with $\lambda$ and $\phi^\omega_\pm$ being replaced by
$-\lambda$ and $\phi^\omega_\mp$.
Combining these results, we obtain
\begin{eqnarray}
  \sum_{g=\pm}\langle x_4;g| \gamma_5e^{-t D^\dagger D}| x_4';g\rangle
  &=& {\rm sign} \lambda \left[\frac{e^{-\lambda^2 t}}{\sqrt{\pi t}}e^{-\frac{(x_4+x_4')^2}{4t}}
    \right.\nonumber\\&&\left.-|\lambda| e^{-|\lambda|(x_4+x_4')}
  \left\{\mbox{erfc}\left(-\frac{x_4+x_4'}{2\sqrt{t}}+|\lambda| \sqrt{t}\right)-2\right\}\right].
 \nonumber\\
\end{eqnarray}

Now we are ready to compute the index by taking trace over
$x$ and $\lambda$,
\begin{eqnarray}
  \label{eq:trace-anti-APS}
  {\rm Tr}\gamma_5e^{-t D^\dagger D}  
  &=& \sum_\lambda {\rm sign} \lambda \int dx_4 \left[\frac{\partial}{\partial x_4}\left\{ \frac{1}{2}e^{-2|\lambda|x_4}
    \mbox{erfc}\left(-\frac{x_4}{\sqrt{t}}+|\lambda| \sqrt{t}\right)\right\}+2|\lambda|e^{-2|\lambda|x_4}\right]
  \nonumber\\
  &=& -\sum_\lambda \frac{{\rm sign}\lambda}{2}\mbox{erfc}\left(|\lambda| \sqrt{t}\right) + \sum_\lambda {\rm sign}\lambda.
\end{eqnarray}
  In the $t\to 0$ limit, the above formula apparently converges to
\begin{eqnarray}
    \lim_{t\to 0}{\rm Tr}\gamma_5e^{-t D^\dagger D}= \sum_\lambda \frac{{\rm sign}\lambda}{2} = \frac{\eta(iD^{\rm 3D})}{2},
\end{eqnarray}
which has the opposite sign to the APS case in Eq.~(\ref{eq:etaflatAPS}). 
We should, however, note that the two terms in Eq.~(\ref{eq:trace-anti-APS}) have different origins.
The first term is a contribution from the nonzero bulk modes,
which is exactly the same as the APS boundary case.
But the second contribution is from the edge-localized zero energy modes,
which cannot be regularized by the exponential factor $e^{-t D^\dagger D}$.
For this reason, the anti-APS boundary case is not appropriate for deriving the index theorem,
since the simple heat-kernel-type regularization is not enough to regulate these edge-localized modes.

\subsection{General gauge background}

The APS index theorem applies only to a compact manifold.
Therefore, the infinite flat cylinder computation in the previous sections
is not complete, as clearly seen by the fact that
$\eta(iD^{\rm 3D})/2$ alone cannot be an integer in general.
On a compact manifold, the ``flatness'' in the $x_4$ direction must be lost
to make the system compactified, otherwise, we need another boundary,
which cancels (the noninteger part of) the eta invariant.
The original APS index theorem \cite{Atiyah:1975jf}
was completed by 
introducing ``doubling'' of a non-flat compact manifold $X$ to eliminate the boundary and
form a closed manifold, and then interpolating the solutions of flat cylinder
and those on the doubled $X$.
Here they still assumed a flatness near the boundary,
so that the flat cylinder solutions well approximate the full ones.

Let us here review the derivation by Alvarez-Gaum\'e {\it et al.} \cite{AlvarezGaume:1984nf} who
introduced two boundaries at $t=-\infty$ and $+\infty$, so that the flat metric is allowed,
and consider a non-trivial $x_4$ dependence of the gauge
fields (here we take $U(1)$ or $SU(N)$ gauge group)
between them to derive the index.

First, the Dirac operator is expressed as
\begin{eqnarray}
D = \tau_1 \otimes 1_{2\times 2}\frac{\partial}{\partial t}-i\tau_2 \otimes H_t,
\end{eqnarray}
where $\tau_i$ denote the Pauli matrices, and $H_t=iD^{\rm 3D}(x_4=t)$.
In the adiabatic approximation, where $H_t$ changes slowly with $t$,
the zero mode solution of $D$ is given by
\begin{eqnarray}
D\psi &=&0,\;\;\; \psi = f(t)\otimes \psi_t, \\
H_t \psi_t &=& \lambda(t)\psi_t,\;\;\; \tau_1(\partial_t +\tau_3 \lambda(t))f(t)=0,\\
f(t)&=& \exp\left(-\int^t dt' \tau_3 \lambda(t')\right)\chi,
\end{eqnarray}
where $\chi$ is a constant.
For the positive chiral mode (here $\tau_3=+1$)
$f(t)$ is normalizable only when $\lambda(-\infty)<0$
and $\lambda(+\infty)>0$, while the negative chiral mode has opposite signs of the eigenvalue.
In either case, $\lambda(t)$ changes its sign somewhere in the $t$ history.
Namely, the APS index counts the zero-crossings of the eigenvalues of $H_t$,
which can be expressed by
\begin{eqnarray}
\mathcal{I} = \frac{1}{2}\left[\eta(H_{t=+\infty})-\eta(H_{t=-\infty})\right]-\frac{1}{2}\int^\infty_{-\infty} dt' \frac{d}{dt'}\eta(H_{t'}),
\end{eqnarray}
where the second term is necessary to cancel the non-integer part of the eta invariants.

The remaining task is to show that the second term is equivalent to
the four-dimensional integral of the conventional instanton density.
To this end, we first express the eta-invariant in integral-form
\begin{eqnarray}
\eta(H_t) = \lim_{s\to 0}\frac{2}{\Gamma(\frac{s+1}{2})}\int_0^\infty du u^s {\rm Tr}H_t e^{-u^2H_t^2},
\end{eqnarray}
and compute its $t$ derivative,
\begin{eqnarray}
\label{eq:etaderivative}
-\frac{1}{2}\frac{d}{dt}\eta(H_t) &=& -\lim_{s\to 0}\frac{1}{\Gamma(\frac{s+1}{2})}\int_0^\infty du u^s \frac{\partial}{\partial u}
{\rm Tr}\left[u\frac{\partial H_t}{\partial t} e^{-u^2H_t^2}\right]
\nonumber\\
&=&\frac{1}{\sqrt{\pi}}\lim_{u\to 0}{\rm Tr}\left[u\frac{\partial H_t}{\partial t} e^{-u^2H_t^2}\right],
\end{eqnarray}
where we have taken the $s\to 0$ limit and the trace ${\rm Tr}$ is taken over two-component spinor,
color, and three-dimensional coordinates.
Then one can relate this quantity to the three-dimensional integral of the instanton density at $x_4=t$ by
\begin{eqnarray}
\int d^3 x \epsilon_{\mu\nu\rho\sigma}{\rm tr}_cF^{\mu\nu}F^{\rho\sigma}(x_4=t)
&=& \lim_{u\to 0} \int d^3x\; {\rm tr}_{c,s}\gamma_5e^{u^2D^2}
\nonumber\\
&=& \lim_{u\to 0} {\rm Tr}\;{\rm tr}_{s'}\tau_3 e^{u^2 (\partial_t^2-H_t^2+\tau_3 \frac{\partial H_t}{\partial t})}
\nonumber\\
&=& \lim_{u\to 0} \int_{-\infty}^{\infty}\frac{d\omega}{2\pi}
{\rm Tr}\;{\rm tr}_{s'}\tau_3 e^{u^2 ((i\omega+\partial_t)^2-H_t^2+\tau_3 \frac{\partial H_t}{\partial t})}
\nonumber\\
&=& \lim_{u\to 0} \int_{-\infty}^{\infty}\frac{d\omega}{2\pi}e^{-\omega^2 u^2}
{\rm Tr}\;{\rm tr}_{s'}\tau_3 \left(u^2\tau_3 \frac{\partial H_t}{\partial t}\right)e^{-u^2 H_t^2}
\nonumber\\
&=& \frac{1}{\sqrt{\pi}}\lim_{u\to 0}{\rm Tr}\left[u\frac{\partial H_t}{\partial t} e^{-u^2H_t^2}\right],
\end{eqnarray}
which agrees with Eq.~(\ref{eq:etaderivative}).
Identifying $t=x_4$ and $H_t= -i D^{\rm 3D}(x_4=t)$, 
we obtain
\begin{eqnarray}
\mathcal{I}= \int d^4x\;
\epsilon_{\mu\nu\rho\sigma}{\rm tr}_cF^{\mu\nu}F^{\rho\sigma}
-\frac{1}{2}\left[\eta(iD^{\rm 3D}(-\infty))-\eta(iD^{\rm 3D}(+\infty))\right].
\end{eqnarray}

It is important to note that the above result is obtained by
the standard Fujikawa method: inserting the conventional
plane wave solutions in the $x_4$ direction.
This is valid only when $t$ dependence is negligible
at the boundaries $t=\pm \infty$.
Namely, this computation is done in a set-up where
the role of edge modes is not relevant.
The interactions between edge and bulk modes are turned off.
For the more general $x_4$ dependent gauge background,
the standard Fujikawa method is difficult since
the APS boundary condition requires non-perturbative
information of the eigenfunctions of $D^{\rm 3D}$.
As is discussed later, the APS boundary condition
has more fundamental problems in application to the physical fermion system with boundaries.

\subsection{Difference between the $\eta$ invariant and the Chern-Simons term}

So far, the $\eta$ invariant has been defined by the eigenvalues of the
Dirac operator on the surface, and it has not been shown 
how it is perturbatively expressed. 
It is known that $\eta(iD^{\rm 3D})$ 
formally appears in the phase of a ``massive'' Dirac fermion determinant,
\begin{eqnarray}
\det \frac{D^{\rm 3D}-M}{D^{\rm 3D}+\Lambda} \propto
\exp\left[i\pi\eta(iD^{\rm 3D})\right],
\end{eqnarray}
where we have introduced the Pauli-Villars regulator,
assuming both $M$ and $\Lambda$ are positive (and large),
and it is perturbatively equivalent to
\begin{eqnarray}
\exp(iCS),
\end{eqnarray}
which can be obtained from an integral 
\begin{eqnarray}
CS &=& \int_0^u du \frac{d}{du} {\rm Im}
\ln \det \frac{D^{\rm 3D}(u)-M}{D^{\rm 3D}(u)+\Lambda},
\end{eqnarray}
up to $1/M$ and $1/\Lambda$ corrections.
Here $D^{\rm 3D}(u)$ is the Dirac operator
with $uA_\mu$, which denotes a linear one-parameter
deformation of the original gauge field.

In the above massive fermion determinant, it is no problem
to identify the Chern-Simons action $CS$ with the $\eta$ invariant.
However, in the index theorem, they are different,
since $CS$ is not gauge invariant under a ``large'' gauge
transformation with a winding number $n$,
\begin{eqnarray}
CS \to CS + 2\pi n.
\end{eqnarray}
Since $\eta(iD^{\rm 3D})/2$ should be obtained in a gauge invariant
regularization, it differs from $CS/2\pi$ by an integer,
which is not gauge invariant.

In Appendix~\ref{app:2dexample}, we exactly compute
the $\eta$ invariant in one-dimensional QED with flat background field
and obtain
\begin{eqnarray}
\label{eq:etaandCS}
\frac{\eta}{2} = \frac{CS}{2\pi}-\left[\frac{CS}{2\pi}\right],
\end{eqnarray}
(up to an irrelevant constant) where $[f]$
denotes the Gauss symbol or the greatest integer less than or equal to $f$.
Although we have not found any proof in the literature,
we assume that this expression is generally valid in the following discussions\footnote{
Our argument cannot exclude a possibility of an additional gauge invariant integer,
which is non-locally given.
},
even for the three-dimensional case with non-Abelian gauge fields.
In fact, Eq.~(\ref{eq:etaandCS}) has good properties listed below.
It is 1) manifestly gauge invariant,
2) reflects non-locality of the APS boundary condition as
the Gauss symbol is highly nonlocal,
3) shows that the total APS index is no longer
a topological invariant, since $\eta/2$ can discretely jump
by an integer\footnote{This jump is induced by the level crossing of the
surface Dirac operator.}, and
4) shows non-compatibility of the gauge invariance and
the T (or parity) invariance of the massless Dirac fermion determinant.
To confirm the last property, let us consider the massless fermion
determinant with Pauli-Villars regulator,
\begin{eqnarray}
\det \frac{D^{\rm 3D}}{D^{\rm 3D}+\Lambda} \propto
\exp\left[i\pi\eta(iD^{\rm 3D})/2\right],
\end{eqnarray}
which is gauge invariant but breaks the T invariance.
To recover the T invariance, the only possible local counterterm we can add
is $\exp(-iCS/2)$; then the remaining phase $\exp(-i\pi[CS/2\pi])$
breaks the gauge invariance \cite{Redlich:1983dv}, 
by the same mechanism as Witten's
global anomaly \cite{Witten:1982fp}.

\subsection{APS boundary condition unlikely to be realized in physics}

The APS boundary condition commutes with $\gamma_5$,
and, therefore, preserves helicity.
Namely this boundary condition keeps the fermion, on which the
Dirac operator operates, massless \cite{Hortacsu:1980kv}.
This looks like a reasonable choice but when we consider
reflection of the fermions at the boundary,
we find that the APS boundary condition is very unnatural\footnote{
One may consider a possibility that fermions
are never reflected, which is another unnatural set-up,
where energy that fermion carries is accumulated
at the boundary and never goes back to the bulk.
}.

Consider a flat surface of some material at $x_4=0$.
Unless the boundary fermion is somehow polarized,
for example, by an anisotropic crystal structure,
it is natural to assume that the system is rotationally symmetric
along the $x_4$ axis perpendicular to the surface.
This $SO(3)$ (or $SO(2,1)$ in Minkowski space-time) rotational symmetry
is essential for the edge-localized mode of topological insulators
to act as a relativistic Dirac fermion.
However, this $SO(3)$ symmetry is not compatible with
the helicity conservation,
which is respected by the APS boundary condition, because it
requires a spin flip whenever fermions are reflected at the boundary \cite{Luscher:2006df}.

In mathematics, we can impose any boundary condition on the first order
differential equations.
However, this is not true in quantum field theory, where we need to regularize
them by subtraction equations and take the continuum limit.
In the lattice gauge theory, it is known that any local boundary condition
except for the Dirichlet boundary requires a fine tuning or some additional
symmetry on the boundary to protect it in the continuum limit.
As the quantum field theory is formulated to somehow neglect 
short-range structure, this requirement of fine-tuning
should be universal for any regularization.
Since the APS boundary condition is nonlocal, this argument
cannot be directly applied, but it is unlikely that
the nonlocality helps to make the boundary condition stable.
Therefore, we conclude that the APS boundary condition is
unlikely to be realized in the physical fermion system with boundary.

Treating a manifold with boundary as a closed system is
also unnatural in physics, as any boundary in our world has ``outside'' of it.
The surface of the topological insulator is nontrivial because
its outside is not empty but surrounded by normal insulator.

The above discussion suggests to us a need to consider
more natural set-up in physics.
We should have a domain-wall, like the one between topological and normal
insulators, rather than a simple boundary without an outside.
It is more natural to have a massive fermion since it is not the helicity
but rotational symmetry that should be preserved.
It is better to have a boundary condition not imposed by hand but
automatically and locally given by dynamics of the system.

Can we still define an index for such a massive Dirac operator?
As is shown in the next section, the answer is ``yes''.
We introduce the so-called domain-wall Dirac fermion operator,
which is massive in the bulk and gapless at the boundary.
Its local boundary condition is not imposed by hand but automatically
satisfied by the kink structure of the mass term.
Therefore, no fine-tuning is needed.
We define an index by its eta-invariant, to which
the edge-localized gapless modes play a crucial role.
Moreover, the new index coincides with the original APS index.



\section{APS index from domain-wall fermion Dirac operator}
\label{sec:domain-wall}

In this section, we consider a different setup from the original work
by APS \cite{Atiyah:1975jf}.
So far we have considered a manifold with boundary, as a closed system.
But in real physics, no boundary can exist without ``outside'' of the region.
For example, the boundary of the topological insulator
is always surrounded by the normal insulator.
We cannot say on which the edge-localized modes reside,
since they require both sides to support them,
unless the gap is infinitely large.

In this respect, the so-called domain-wall fermion \cite{Callan:1984sa, Kaplan:1992bt}
is a more appropriate
setup for the physical system with boundary.
The domain-wall fermion Dirac operator is defined by
\begin{eqnarray}
  D_{DW} = D + M\epsilon(x_4),\;\;\epsilon(x_4) = {\rm sign}\; x_4.
\end{eqnarray}
where the mass term flips its sign across the domain-wall located at $x_4=0$.
Here and in the following, we take $M$ to be positive.
In lattice gauge theory, we often consider the domain-wall fermion determinant
together with a Pauli-Villars field,
\begin{eqnarray}
  \label{eq:DWdet}
  \det \frac{D + M\epsilon(x_4)}{D-M},
\end{eqnarray}
to cancel the bulk mode effects in the region $x_4<0$.
Note here that fermion field is defined in the
whole $-\infty<x_4<\infty$ region and
no boundary condition is imposed on it\footnote{
  Strictly speaking, we should give an IR cutoff by compactifying the
  manifold with some appropriate boundary condition,
  such as periodic boundary condition.
  Then we need another domain-wall at some point of $x_4$.
  We will discuss this anti-domain-wall fermion contribution
  at the end of this section. 
  }.
Therefore, this determinant provides a good model to describe
fermions in a topological insulator located in the $x_4>0$ region,
surrounded by a normal insulator sitting in the $x_4<0$ region.
As we explicitly show, the edge-localized modes appear at the boundary $x_4=0$,
and play a crucial role in the definition of the index.

The determinant Eq.~(\ref{eq:DWdet}) 
is real\footnote{
This is true even with  a naive lattice regularization using the  Wilson Dirac operator.
},  due to the ``$\gamma_5$ Hermiticity'',
\begin{eqnarray}
\label{eq:DWdet2}
  \det \left[(D + M\epsilon(x_4))(D - M)^{-1}\right]
  &=& \det \left[\gamma_5^2(D + M\epsilon(x_4))\gamma_5^2(D - M)^{-1}\right]\nonumber\\
  &=& \det \left[(D^\dagger + M\epsilon(x_4))(D^\dagger - M)^{-1}\right] \nonumber\\
  &=& \left|\det \left[(D^\dagger + M\epsilon(x_4))(D^\dagger - M)^{-1}\right]\right| (-1)^{\mathcal{I}},
\end{eqnarray}
where $\mathcal{I}$ is an integer determining the sign of the determinant.
In fact, we will explicitly show that this integer $\mathcal{I}$
is equivalent to the APS index.
A similar statement is found in \cite{Witten:2015aba},
but neither the explicit bulk fermion determinant
nor its boundary condition is given.
The outside of our target domain is not mentioned, either.
As is shown below, we need neither the massless Dirac operator nor
nonlocal APS boundary condition for the new index.

Our new index $\mathcal{I}$ is formally defined by
a regularized eta invariant of the Hermitian operator
$H_{DW} = \gamma_5 (D + M\epsilon(x_4))$:
\begin{eqnarray}
\mathcal{I} &\equiv& \frac{\eta(H^{reg}_{DW})}{2}
= \frac{1}{2}\eta(H_{DW})- \frac{1}{2}\eta(H_{PV}),
\end{eqnarray}
where we employ the Pauli-Villars regularization
with another Hermitian operator $H_{PV} = \gamma_5 (D-M)$.
This definition coincides with the exponent appearing
in Eq.(\ref{eq:DWdet2}) as 
\begin{eqnarray}
  \det \frac{D + M\epsilon(x_4)}{D-M} &=&
  \det \frac{iH_{DW}}{iH_{PV}}
  =
  \prod_{\lambda_{DW}} i\lambda_{DW} / \prod_{\lambda_{PV}} i\lambda_{PV}
  \nonumber\\
  &\propto& \exp\left(\frac{i\pi}{2}\left(\sum_{\lambda_{DW}}{\rm sign} \lambda_{DW}
  -\sum_{\lambda_{PV}}{\rm sign} \lambda_{PV}\right)\right) \
  \nonumber\\
  &=& (-1)^{\frac{1}{2}\eta(H_{DW})- \frac{1}{2}\eta(H_{PV})}.
\end{eqnarray}
In the following, we compute the two eta invariants
$\eta(H_{DW})$ and $\eta(H_{PV})$ separately, by
introducing another regularization using the (generalized) $\zeta$ function
(we simply call it the $\zeta$-function regularization).
This double regularization is not theoretically needed but
simplifies the computation and clarifies the role of
the Pauli-Villars fields.
In fact, we see that $\eta(H_{DW})/2$ alone gives
only a ``half'' of the (bulk contribution of) total APS index,
to which another ``half'' is provided by $\eta(H_{PV})/2$.

Let us compute $\eta(H_{PV})$ first. 
Interestingly, it coincides with the AS index,
\begin{eqnarray}
  \label{eq:massiveFujikawaclosed}
  \eta(H_{PV}) &=& \lim_{s\to 0}{\rm Tr}\frac{H_{PV}}{(\sqrt{H_{PV}^2})^{1+s}}
  = \lim_{s\to 0}\frac{1}{\Gamma\left(\frac{1+s}{2}\right)}
  \int_0^\infty dt t^{\frac{s-1}{2}}{\rm Tr}H_{PV} e^{-t H_{PV}^2}\nonumber\\
  &=& -\frac{1}{\sqrt{\pi}}\int_0^\infty dt' t'^{-\frac{1}{2}}{\rm Tr} \gamma_5\left(1-\frac{D}{M}\right) e^{-t' D^\dagger D/M^2 }e^{-t'},
  \nonumber\\&=& -\frac{1}{32\pi^2} \int d^4x\; \epsilon_{\mu\nu\rho\sigma}{\rm tr}_cF^{\mu\nu}F^{\rho\sigma} +\mathcal{O}(1/M^2).
\end{eqnarray}
Here we have changed the valuable $t=t'/M^2$,
and the conventional Fujikawa method has been applied to evaluate
${\rm Tr}\gamma_5 e^{-t' D^\dagger D/M^2 }$.
Moreover, we can show that $\eta(H_{PV})$ is independent of $M$ as follows.
Since $\{H_{PV},D\}=0$, every eigenmode
$\phi_{\lambda_{PV}}$ with eigenvalue $\lambda_{PV}$ makes a pair
with $D\phi_{\lambda}$ whose eigenvalue has the opposite
sign $-\lambda_{PV}$, unless $D\phi_{\lambda}=0$.
The zero modes of $D$, which commute with $\gamma_5$,
are simultaneously the eigenmodes of $H_{PV}$, whose
eigenvalues are $\pm M$ with $\gamma_5 = \mp 1$.
Therefore, the left-hand side of Eq.~(\ref{eq:massiveFujikawaclosed})
becomes
\begin{eqnarray}
  \eta(H_{PV}) = - {\rm Tr}_{zeros} \frac{\gamma_5M}{M} = -(n_+-n_-),
\end{eqnarray}
which is independent of $M$.
Here ${\rm Tr}_{zeros}$ is the trace over the zero modes of $D$ only.

It is also interesting to note that the structure of the eta invariant
is naturally embedded in the index theorem of the
massless lattice Dirac operator.
The Neuberger's lattice overlap Dirac operator \cite{Neuberger:1997fp, Neuberger:1998wv}
with the lattice spacing $a$ is defined by
\begin{eqnarray}
  D^{ov} = \frac{1}{a}\left[1 + \gamma_5 \frac{H_W}{\sqrt{H_W^2}}\right],
\end{eqnarray}
where $H_W = \gamma_5(D_W-1/a)$ is the Hermitian Wilson Dirac operator
with the cut-off scale mass $M=-1/a$.
Actually, its index is given by
\begin{eqnarray}
  {\rm Tr}\gamma_5\left(1-\frac{D_{ov}a}{2}\right)
  = -\frac{1}{2}{\rm Tr}\frac{H_W}{\sqrt{H_W^2}} = -\frac{1}{2}\eta(H_W),
\end{eqnarray}
where we have used ${\rm Tr}\gamma_5=0$ with the finite cut-off.
The sign is not important here: it is just a convention of the sign for the mass
compared to the Wilson term, but the factor $1/2$ has a crucial role
to cancel the contribution from the doublers, which plays the same role
of (another) Pauli-Villars field in the continuum.
This computation clearly shows that there is no need to introduce
massless Dirac operator to define the index, at least, on a closed manifold.
As we will show below, this is true even with boundary.

Now our goal in this section is to compute the remaining
contribution $\eta(H_{DW})$ and show 
\begin{eqnarray}
  \eta(H_{DW}) &=&  
  \frac{1}{32\pi^2}\int d^4x\; \epsilon(x_4) \epsilon_{\mu\nu\rho\sigma}{\rm tr}_cF^{\mu\nu}F^{\rho\sigma}
  -\eta(iD^{\rm 3D}).
\end{eqnarray}
For this massive case, we switch to the Dirac representation for
the gamma matrices:
\begin{eqnarray}
  \gamma_{i=1,2,3}&=&\left(\begin{array}{cc}
     & \sigma_i\\
   \sigma_i & 
  \end{array}\right)=\tau_1\otimes \sigma_i,\;\;\;
  \gamma_4=\left(\begin{array}{cc}
    1_{2\times 2} &\\
   & -1_{2\times 2}
  \end{array}\right)=\tau_3\otimes 1_{2\times 2},
  \nonumber\\
  \gamma_5&=&-\gamma_1\gamma_2\gamma_3\gamma_4=\left(\begin{array}{cc}
    & i 1_{2\times 2} \\
   -i 1_{2\times 2} & 
  \end{array}\right)=-\tau_2\otimes 1_{2\times 2}.  
\end{eqnarray}
Our Hermitian Dirac operator is then expressed by
\begin{eqnarray}
  H_{DW} &=&\gamma_5\gamma_4(\partial_4+\gamma_4 M\epsilon(x_4))+B
  \nonumber\\&=& \left(\begin{array}{cc}
     & -i(\partial_4-M\epsilon(x_4)) \\ 
    -i(\partial_4+M\epsilon(x_4))&   
  \end{array}\right)
  +\left(\begin{array}{cc}
    -iD^{\rm 3D} &\\
   & iD^{\rm 3D}
  \end{array}\right),
\end{eqnarray}
where $B=\gamma_5\sum_{i=1}^3 \gamma_i D^i$, and $D^{\rm 3D}=-\sigma_iD^i$.


\subsection{$x_4$-independent background}

As we have demonstrated in the case of the APS boundary,
let us begin with the flat background 
with no $x_4$ dependence of the gauge fields.
Our cylinder is  now extended to the $x_4< 0$ region.
Since $F_{4k}=0$ for any $k$, our goal here is to show
\begin{eqnarray}
  \label{eq:DWcylinder}
  \eta(H_{DW})  =  -\eta(iD^{\rm 3D}).
\end{eqnarray}

With the $A_4=0$ gauge, $H_{DW}^2$ can be written as
\begin{eqnarray}
  H_{DW}^2 = -\partial_4^2 + B^2 + M^2 - 2M\gamma_4\delta(x_4),
\end{eqnarray}
which commutes with $\gamma_4$, and $B$.
It is, therefore, convenient to consider the eigenvalue problem of
$H_{DW}^2$ by assuming the form of the solution as $\varphi_\pm (x_4)\otimes\phi_\lambda^{\rm 3D}(\vec{x})$
where $\varphi_\pm (x_4)$ satisfies
\begin{eqnarray}
  \label{eq:EOM2DW}
  (-\partial_4^2 + \lambda^2+M^2\mp 2M\delta(x_4))\varphi_\pm (x_4)=\Lambda^2 \varphi_\pm (x_4),
\end{eqnarray}
$\phi_\lambda^{\rm 3D}(\vec{x})$ is the eigenfunction of $iD^{\rm 3D}$ with the eigenvalue $\lambda$,
and $\tau_3\varphi_\pm (x_4)=\pm \varphi_\pm (x_4)$.
Note that the eigenvalue of $\tau_3$ corresponds to that of $\gamma_4$.

The solutions to Eq.~(\ref{eq:EOM2DW}) are obtained as
\begin{eqnarray}
  \varphi^\omega_{\pm,o} (x_4) &=& \frac{u_\pm}{\sqrt{4\pi}}\left(e^{i\omega x_4}-e^{-i\omega x_4}\right),\nonumber\\
  \varphi^\omega_{\pm,e} (x_4) &=& \frac{u_\pm}{\sqrt{4\pi(\omega^2+M^2)}}
  \left\{(i\omega\mp M)e^{i\omega |x_4|}+(i\omega\pm M)e^{-i\omega |x_4|}\right\},\nonumber\\
  \varphi^{\rm edge}_{+,e} (x_4) &=& u_+\sqrt{M}e^{-M|x_4|},
\end{eqnarray}
where $\omega=\sqrt{\Lambda^2-\lambda^2-M^2}$, and the subscripts $e,o$ denote even and odd components
under the time reversal $T:x_4\leftrightarrow -x_4$.

We emphasize here that we have not imposed any boundary condition by hand,
but the delta-function potential automatically chooses
non-trivial boundary conditions on the fermion fields:
\begin{eqnarray}
  \left.\left[\frac{\partial}{\partial x_4} \pm M\epsilon(x_4)\right]\varphi^{\omega,\rm edge}_{\pm, e} (x_4)\right|_{x_4=0} = 0,\;\;\;\;\varphi^\omega_{\pm, o} (x_4=0) =0.
\end{eqnarray}
More importantly, these boundary conditions respect the $SO(3)$ rotational
symmetry on the $x_4=0$ surface, rather than helicity.

The above solutions satisfy
\begin{eqnarray}
  \int_{-\infty}^\infty dx_4 [\varphi^{\omega'}_{\pm,e/o}(x_4)]^\dagger\varphi^{\omega}_{\pm,e/o}(x_4) &=& \delta(\omega-\omega'),\\
  \int_{-\infty}^\infty dx_4 [\varphi^{\rm edge}_{+,e} (x_4)]^\dagger\varphi^{\rm edge}_{+,e} (x_4) &=& 1,
\end{eqnarray}
for positive $\omega$ and $\omega'$. They also satisfy the completeness condition
in a subspace where $iD^{3D}$ takes the eigenvalue $\lambda$, for which
$B$ takes the eigenvalue $\mp \lambda$ for $\gamma_4=\pm 1$ eigenmodes,
\begin{eqnarray}
  \label{eq:completesetDW}
\sum_{a=e,o} \int_0^\infty d\omega [\varphi^\omega_{+,a} (x_4)] [\varphi^\omega_{+,a} (x_4')]^\dagger
+[\varphi^{\rm edge}_{+,e} (x_4)] [\varphi^{\rm edge}_{+,e} (x_4')]^\dagger
= \delta(x_4-x_4')1_{2\times 2},
\nonumber\\
\int_0^\infty d\omega [\varphi^\omega_{-,a} (x_4)] [\varphi^\omega_{-,a} (x_4')]^\dagger
= \delta(x_4-x_4')1_{2\times 2},\;
\end{eqnarray}
for any $x_4,x_4'$.

Next, let us compute the kernel of the operator $H_{DW} e^{-t H_{DW}^2}$
using the complete set obtained above for each $\lambda$.
For the $++$ component, we have
\begin{eqnarray}
  \langle x_4;+| H_{DW} e^{-t H_{DW}^2}| x_4';+\rangle
  &=&-\lambda
\left\{\sum_{a=e,o}\int_0^\infty d\omega e^{-t(\omega^2+\lambda^2+M^2)}
    [\varphi^\omega_{+,a} (x_4)]_+ [\varphi^\omega_{+,a} (x_4')]_+^*
    \right.\nonumber\\&&\left.+e^{-\lambda^2 t}[\varphi^{\rm edge}_{+,e} (x_4)]_+ [\varphi^{\rm edge}_{+,e} (x_4')]_+^*\right\}
    \nonumber\\&=&
-\lambda \frac{e^{-(\lambda^2+M^2) t}}{\sqrt{4\pi t}}e^{-\frac{(x_4-x_4')^2}{4t}}
  \nonumber\\&&
  +\frac{\lambda M}{2} e^{-\lambda^2 t} e^{-M(|x_4|+|x_4'|)}\mbox{erfc}\left(-\frac{|x_4|+|x_4'|}{2\sqrt{t}}+M \sqrt{t}\right)
  \nonumber\\&&
  -\lambda M e^{-\lambda^2 t}e^{-M(|x_4|+|x_4'|)}, 
\end{eqnarray}
where we have used $f((|x_4|+|x_4'|)^2)+f((|x_4|-|x_4'|)^2)=f((x_4+x_4')^2)+f((x_4-x_4')^2)$
for any function $f(x)$.
Note that the third term is the contribution from the edge-mode.
The $--$ component is similarly obtained as
\begin{eqnarray}
  \langle x_4;-| H_{DW} e^{-t H_{DW}^2}| x_4';-\rangle
  &=& \lambda \frac{e^{-(\lambda^2+M^2) t}}{\sqrt{4\pi t}}e^{-\frac{(x_4-x_4')^2}{4t}}
  \nonumber\\&&- \frac{\lambda M}{2} e^{-\lambda^2 t}e^{M(|x_4|+|x_4'|)}
  \mbox{erfc}\left(\frac{|x_4|+|x_4'|}{2\sqrt{t}}+M \sqrt{t}\right).
\end{eqnarray}
The trace is given by 
\begin{eqnarray}
  {\rm Tr} H_{DW} e^{-t H_{DW}^2} 
  &=& \frac{1}{2}\sum_\lambda \lambda e^{-\lambda^2 t} \int_{-\infty}^\infty d x_4
  \left[M e^{-2M|x_4|}
    \left\{\mbox{erfc}\left(-\frac{|x_4|}{\sqrt{t}}+M \sqrt{t}\right)-2\right\}\right]  \nonumber\\&&-
  \frac{1}{2}\sum_\lambda \lambda e^{-\lambda^2 t} \int_{-\infty}^\infty d x_4
  \left[M e^{2M|x_4|}
    \mbox{erfc}\left(\frac{|x_4|}{\sqrt{t}}+M \sqrt{t}\right)\right]  
\nonumber\\&=&
    \frac{1}{2}\sum_\lambda \lambda e^{-\lambda^2 t} 
    \left[2\mbox{erfc}\left(M \sqrt{t}\right)-2\right].
\end{eqnarray}
The first term in the parenthesis with the complementary error function
is from the bulk mode, while the second term is from the edge mode.

In the $M\to \infty$ limit, $\mbox{erfc}\left(M \sqrt{t}\right)$ vanishes
and we obtain the desired result,
\begin{eqnarray}
 \label{eq:APSindexDWflat}
  \eta(H_{DW}) &=& -\sum_\lambda \frac{\lambda}{\sqrt{\pi}}
  \int_0^\infty dt t^{-1/2}  e^{-\lambda^2 t} =
  - \sum_\lambda \frac{\lambda}{|\lambda|} = -\eta(iD^{\rm 3D}).
\end{eqnarray}
It is clear that the $\eta(iD^{\rm 3D})$ comes entirely
from the edge-localized modes.



\subsection{Fujikawa method for general background}
\label{subsec:DWFujikawa}

Let us consider the general gauge field background and complete
the APS index theorem using the Fujikawa method.

Here we can keep the $A_4=0$ gauge, and therefore,
the wave functions in the $x_4$ direction in the previous subsection
$\varphi_{\pm,e/o}^{\omega}$ and $\varphi_{+,e}^{\rm edge}$ are still useful.
Thus, we only need to replace the three-dimensional part of the wave function,
which was given by the eigenfunction $\phi_\lambda^{\rm 3D}$ of $iD^{\rm 3D}$,
by that of the plane wave,
\begin{eqnarray}
  \phi_{\bm{p},\uparrow\downarrow}^{\rm 3D}(\bm{x}) = \frac{v_{\uparrow\downarrow}}{(2\pi)^{3/2}}e^{i\bm{p}\cdot \bm{x}},
\end{eqnarray}
where $\bm{p}=(p_1,p_2,p_3)$ and $\bm{x}=(x_1,x_2,x_3)$ are the spatial
components of momentum and position, respectively\footnote{
Here we assume that the spatial directions are infinitely large for simplicity.
}.
The spin degrees of freedom are described by
\begin{eqnarray}
  v_\uparrow= \left(
  \begin{array}{c}
      1\\
      0
  \end{array}
  \right),\;\;\;
   v_\downarrow= \left(
  \begin{array}{c}
      0\\
      1
  \end{array}
  \right).
\end{eqnarray}

Let us here summarize what we will compute in this subsection.
Our goal is to obtain the index for general gauge background,
defined by the eta invariant,
\begin{eqnarray}
  \eta(H_{DW}) = \lim_{s\to 0}\left[{\rm Tr}(M\gamma_5\epsilon(x_4))\left(\sqrt{H_{DW}^2}\right)^{-1-s}
  +{\rm Tr}(\gamma_5 D)\left(\sqrt{H_{DW}^2}\right)^{-1-s}\right].  
\end{eqnarray}
Since the second term includes contribution from
massless edge-localized modes, it is non-local in general.
Following the general strategy to compute the ``local'' part
of the phase of the odd-dimensional massless fermion determinant \cite{AlvarezGaume:1984nf},
we consider a one-parameter family of gauge fields $uA_\mu$,
and take a $u$-derivative and integrate it again,
\begin{eqnarray}
  \int_0^1 du \frac{d}{du}\left[{\rm Tr}(H_{DW}(u)-M\gamma_5\epsilon(x_4))\left(\sqrt{H_{DW}(u)^2}\right)^{-1-s}\right]
    \nonumber\\
    =\int_0^1 du {\rm Tr}\left[-s\frac{d}{du}H_{DW}(u)\left(\sqrt{H_{DW}(u)^2}\right)^{-1-s}
    -\frac{d}{du}\left(\gamma_5 M\epsilon(x_4)\left(\sqrt{H_{DW}(u)^2}\right)^{-1-s}\right)\right],
\end{eqnarray}
where $H_{DW}(u)$ is the corresponding domain-wall fermion Dirac operator at $u$.
This procedure allows us to compute the eta invariant
up to an integer, which may depend on a winding number of gauge transformation
on the surface.
Using the formula
\begin{eqnarray}
  \frac{1}{(\sqrt{O^2})^{1+s}} =
   \frac{1}{\Gamma\left(\frac{1+s}{2}\right)}\int_0^\infty dt\; t^{\frac{s-1}{2}}e^{-t O^2},
\end{eqnarray}
for a Hermitian operator $O$, our goal is to compute
\begin{eqnarray}
  \label{eq:goalDW}
  \eta(H_{DW}) &=& \lim_{s\to 0}\frac{1}{\Gamma\left(\frac{1+s}{2}\right)}\int_0^\infty dt\; t^{\frac{s-1}{2}}
  \lim_{M\to \infty}{\rm Tr}\left[\gamma_5\epsilon(x_4) e^{-t \frac{H_{DW}^2}{M^2}}\right]
  \nonumber\\&&
  +\int_0^1 du \lim_{s\to 0}\frac{1}{\Gamma\left(\frac{1+s}{2}\right)}\int_0^\infty dt\; t^{\frac{s-1}{2}}
  \lim_{M\to \infty}{\rm Tr}\left[-s\frac{dH_{DW}(u)}{du}\frac{e^{-t \frac{H_{DW}(u)^2}{M^2}}}{M}\right]
  \nonumber\\&&
  -\int_0^1 du \frac{d}{du}\left\{\lim_{s\to 0}\frac{1}{\Gamma\left(\frac{1+s}{2}\right)}\int_0^\infty dt\; t^{\frac{s-1}{2}}
  \lim_{M\to \infty} {\rm Tr}\left[\gamma_5\epsilon(x_4) e^{-t \frac{H_{DW}(u)^2}{M^2}}\right]\right\},
\end{eqnarray}
inserting our complete set $\{\phi_{\bm{p},\uparrow\downarrow}^{\rm 3D}(\bm{x})\otimes\varphi_{\pm,e/o}^{\omega}(x_4)\}$, and
$\{\phi_{\bm{p},\uparrow\downarrow}^{\rm 3D}(\bm{x})\otimes\varphi_{+,e}^{\rm edge}(x_4)\}$ to the trace.
To make $t$ dimensionless, we have rescaled $H_{DW}$ to $H_{DW}/M$.
The third term can be easily evaluated once the first term is obtained.

\subsubsection{The first term of Eq.~(\ref{eq:goalDW})}

Let us evaluate the first term in Eq.~(\ref{eq:goalDW}), using a general formula
\begin{eqnarray}
  {\rm Tr}f\left(\frac{H_{DW}^2}{M^2}\right) &=& \int d^4x
  \sum_{g=\pm}\sum_{a=e,o}\int_0^\infty d\omega \sum_{\sigma=\uparrow\downarrow} \int d^3p \;{\rm tr}_{c}
      [\phi^{\rm 3D}_{\bm{p},\sigma}(\bm{x})\varphi^\omega_{g,a}(x_4)]^\dagger
\nonumber\\&&\times
      f\left(\frac{H_{DW}^2}{M^2}\right)
      [\phi^{\rm 3D}_{\bm{p},\sigma}(\bm{x})\varphi^\omega_{g,a}(x_4)]\nonumber\\
      &&+ \int d^4x
      \sum_{\sigma=\uparrow\downarrow} \int d^3p \;{\rm tr}_{c}
          \left\{[\phi^{\rm 3D}_{\bm{p},\sigma}(\bm{x})\varphi^{\rm edge}_{+,e}(x_4)]^\dagger
          f\left(\frac{H_{DW}^2}{M^2}\right)
      [\phi^{\rm 3D}_{\bm{p},\sigma}(\bm{x})\varphi^{\rm edge}_{+,e}(x_4)]\right\}
      \nonumber\\
      &\hspace{-0in}=& \int d^4x
      \sum_{g=\pm}\sum_{a=e,o}\int_0^\infty d\omega \sum_{\sigma=\uparrow\downarrow} \int \frac{d^3p}{(2\pi)^3} \;{\rm tr}_{c}
      \left\{[v_{\sigma}\varphi^\omega_{g,a}(x_4)]^\dagger
\right.\nonumber\\&&\times
\left.f\left(1-\frac{(ip^i\gamma_i+D)^2+2M\gamma_4\delta(x_4)}{M^2}\right)
  [v_{\sigma}\varphi^\omega_{g,a}(x_4)]\right\}\nonumber\\
      &&+\int d^4x
      \sum_{\sigma=\uparrow\downarrow} \int \frac{d^3p}{(2\pi)^3} \;{\rm tr}_{c}
          \left\{[v_{\sigma}\varphi^{\rm edge}_{+,e}(x_4)]^\dagger
          \right.\nonumber\\&&\times\left.
      f\left(1-\frac{(ip^i\gamma_i+D)^2+2M\gamma_4\delta(x_4)}{M^2}\right)[v_{\sigma}\varphi^{\rm edge}_{+,e}(x_4)]\right\}.
\end{eqnarray}
for any (finite) function $f$.
We can see that in the expansion of $e^{-t \frac{H_{DW}^2}{M^2}}$ in $1/M^2$,
only the term proportional to  $t^2$ and having four different gamma matrices
can contribute to the trace.
Namely, we only need to evaluate
\begin{eqnarray}
  \varphi^\omega_{g,a}(x_4)^\dagger \epsilon(x_4)\left[
    \gamma_5 e^{-t (1-\frac{(ip^i\gamma_i+D)^2+2M\gamma_4\delta(x_4)}{M^2})}\right]
  \varphi^\omega_{g,a}(x_4) &=&
  \epsilon(x_4)\varphi^\omega_{g,a}(x_4)^\dagger  e^{-\frac{t}{M^2}(\omega^2+\bm{p}^2)-t}
  \nonumber\\
&&\hspace{-3.5in}\times  \left[-\frac{t^2}{8M^4}\gamma_5\left\{
    [\gamma_i,\gamma_j]\gamma_4\gamma_k F^{ij}F^{4k} + \gamma_4\gamma_k[\gamma_i,\gamma_j]F^{4k}F^{ij}
    \right\}\right]\varphi^\omega_{g,a}(x_4)
   \nonumber\\
   &&\hspace{-3.5in}= -1_{2\times 2}\epsilon(x_4)\varphi^\omega_{g,a}(x_4)^\dagger \varphi^\omega_{g,a}(x_4) e^{-\frac{t}{M^2}(\omega^2+\bm{p}^2)-t}  
   \frac{t^2}{4M^4}\epsilon_{ijk} (F^{ij}F^{4k} + F^{4k}F^{ij}),
\end{eqnarray}
and similarly,
\begin{eqnarray}
  \varphi^{\rm edge}_{+,e}(x_4)^\dagger \epsilon(x_4)
  \left[\gamma_5e^{-t (1-\frac{(ip^i\gamma_i+D)^2
        +2M\gamma_4\delta(x_4)}{M^2})}\right]
    \varphi^{\rm edge}_{+,e}(x_4)
   \nonumber\\
   &&\hspace{-3.2in}=
   -1_{2\times 2}\epsilon(x_4)\varphi^{\rm edge}_{+,e}(x_4)^\dagger \varphi^{\rm edge}_{+,e}(x_4) e^{-\frac{t}{M^2}(\bm{p}^2)}
   \frac{t^2}{4M^4}\epsilon_{ijk} (F^{ij}F^{4k} + F^{4k}F^{ij}).  
\end{eqnarray}

Then we have
\begin{eqnarray}
  {\rm Tr}\left[\epsilon(x_4)\gamma_5 e^{-t \frac{H_{DW}^2}{M^2}}\right] &=& -\int d^4x\;\epsilon(x_4)
  \;\frac{\sqrt{\pi t}}{8\pi^2}
  \left(\frac{e^{-t}}{\sqrt{\pi t}}-h(t; x_4,M)\right)
  \epsilon_{ijk}{\rm tr}_{c}F^{ij}F^{4k}.
 \end{eqnarray}
Here
\begin{eqnarray} 
  h(t;x_4,M) &=& \frac{1}{2}e^{2M|x_4|}{\rm erfc}\left(\frac{M|x_4|}{\sqrt{t}}+\sqrt{t}\right)
  +\frac{1}{2}e^{-2M|x_4|}\left[{\rm erfc}\left(-\frac{M|x_4|}{\sqrt{t}}+\sqrt{t}\right)-2\right],
\end{eqnarray}
and we have used
\begin{eqnarray}
  \label{eq:spintr}
  \sum_{\sigma=\uparrow\downarrow} v_{\sigma}^\dagger v_{\sigma} &=& 2,
\end{eqnarray}
\begin{eqnarray}
  \label{eq:pint}
  \int \frac{d^3p}{(2\pi)^3} e^{-t\bm{p}^2/M^2} &=& \frac{M^3\sqrt{\pi}}{8\pi^2 t\sqrt{t}},
\end{eqnarray}
and
\begin{eqnarray}
  \sum_{a=e,o}\sum_{g=\pm}\int_0^\infty d\omega  \varphi^\omega_{g,a}(x_4)^\dagger \varphi^\omega_{g,a}(x_4) e^{-\frac{t}{M^2}(\omega^2+M^2)}
  + \varphi^{\rm edge}_{+,e}(x_4)^\dagger \varphi^{\rm edge}_{+,e}(x_4)
  \nonumber\\= M \left[\frac{e^{-t}}{\sqrt{\pi t}}-\frac{1}{2}e^{2M|x_4|}{\rm erfc}\left(\frac{M|x_4|}{\sqrt{t}}+\sqrt{t}\right)
    -\frac{1}{2}e^{-2M|x_4|}{\rm erfc}\left(-\frac{M|x_4|}{\sqrt{t}}+\sqrt{t}\right)+1\right].
\end{eqnarray}

With the $t$ integrals 
\begin{eqnarray}
  \lim_{s\to 0}\frac{1}{\Gamma\left(\frac{1+s}{2}\right)}\int_0^\infty dt\; t^{\frac{s-1}{2}} e^{-t} =1,
\end{eqnarray}
\begin{eqnarray}
  \lim_{s\to 0}\frac{1}{\Gamma\left(\frac{1+s}{2}\right)}\int_0^\infty dt\; t^{\frac{s-1}{2}}
  \sqrt{\pi t}e^{2M|x_4|}{\rm erfc}\left(\frac{M|x_4|}{\sqrt{t}}+\sqrt{t}\right) = \frac{e^{-2 M|x_4|}}{2},
\end{eqnarray}
\begin{eqnarray}
  \lim_{s\to 0}\frac{\sqrt{\pi}}{\Gamma\left(\frac{1+s}{2}\right)}\int_0^\infty dt\; t^{\frac{s}{2}}[e^{-2M|x_4|}]= \lim_{T\to \infty} Te^{-2M|x_4|},
  \end{eqnarray}
  and
\begin{eqnarray}
  \lim_{s\to 0}\frac{\sqrt{\pi}}{\Gamma\left(\frac{1+s}{2}\right)}\int_0^\infty dt\; t^{\frac{s}{2}}
  \left[-e^{-2M|x_4|}{\rm erfc}\left(\frac{-M|x_4|}{\sqrt{t}}+\sqrt{t}\right)\right]
  = - e^{-2M|x_4|}\frac{1+4M |x_4|}{2},
\end{eqnarray}
we have 
\begin{eqnarray}
  g(x_4,M)=\lim_{s\to 0}\frac{1}{\Gamma\left(\frac{1+s}{2}\right)}\int_0^\infty dt\; t^{\frac{s-1}{2}}\sqrt{\pi t}h(t; x_4,M)
  = \frac{e^{-2M|x_4|}}{2}\left(1+2M|x_4|-2\lim_{T\to \infty}T\right),
  \nonumber\\
\end{eqnarray}
with which any finite function $f(x_4)$ gives
\begin{eqnarray}
  \label{eq:edgeinbulk1}
  \lim_{M\to\infty}\int_0^\infty dx_4 g(x_4,M)  f(x_4) &<& \lim_{M\to\infty}\int_0^\infty dx_4 g(x_4,M)  |f^{max}|
  \nonumber\\
  &=&\lim_{T\to \infty}\lim_{M\to\infty}\frac{|f^{max}|(T-1)}{2M} \to 0,
\end{eqnarray}
where $|f^{max}|$ denotes the maximum of $f(x_4)$, and the same is true in the $x_4<0$ region.
Therefore, we obtain the first term in Eq.~(\ref{eq:goalDW}) as
\begin{eqnarray}
\lim_{s\to 0}\frac{1}{\Gamma\left(\frac{1+s}{2}\right)}\int_0^\infty dt\; t^{\frac{s-1}{2}}
\lim_{M\to \infty}{\rm Tr}\left[\gamma_5\epsilon(x_4) e^{-t \frac{H_{DW}^2}{M^2}}\right]
&=& - \int d^4x \epsilon(x_4)\frac{1}{8\pi^2}\epsilon_{ijk}{\rm tr}_{c}F^{ij}F^{4k} 
\nonumber\\
&=&\frac{1}{32\pi^2}\int d^4x\epsilon(x_4)
\epsilon_{\mu\nu\rho\sigma}{\rm tr}_{c}F^{\mu\nu}F^{\rho\sigma}.\nonumber\\
\end{eqnarray}

\subsubsection{The third term of Eq.~(\ref{eq:goalDW})}
Noticing 
\begin{eqnarray}
  \frac{1}{32\pi^2}\int_{x_4>0} d^4x
  \epsilon_{\mu\nu\rho\sigma}{\rm tr}_{c}F^{\mu\nu}F^{\rho\sigma} &=& \frac{1}{2\pi}CS|_{x_4=0} +\mbox{integer},\\ 
   \frac{1}{32\pi^2}\int_{x_4<0} d^4x
  \epsilon_{\mu\nu\rho\sigma}{\rm tr}_{c}F^{\mu\nu}F^{\rho\sigma} &=& -\frac{1}{2\pi}CS|_{x_4=0} +\mbox{integer}, 
\end{eqnarray}
we can compute the third term in Eq.~(\ref{eq:goalDW}) as
\begin{eqnarray}
-\int_0^1 du \frac{d}{du}(\frac{1}{\pi}CS^u|_{x_4=0}) = -\frac{1}{\pi}CS|_{x_4=0}, 
\end{eqnarray}
where $CS^u$ means the Chern-Simons term with the gauge field $uA_\mu$. 

\subsubsection{The second term of Eq.~(\ref{eq:goalDW})}
For the second term in Eq.~(\ref{eq:goalDW}), only the linear term in $t$ in the exponential
$e^{-t \frac{H_{DW}^2}{M^2}}$ with two different gamma matrices in spatial directions
can contribute. Therefore, we need 
\begin{eqnarray}
  \varphi^\omega_{g,a}(x_4)^\dagger \left[-s i \gamma_5\gamma^k A_k e^{-t (1-\frac{(ip^i\gamma_i+D^u)^2+2M\gamma_4\delta(x_4)}{M^2})}\right] \varphi^\omega_{g,a}(x_4) &=&
  \varphi^\omega_{g,a}(x_4)^\dagger  e^{-\frac{t}{M^2}(\omega^2+\bm{p}^2)-t}
  \nonumber\\
&&\hspace{-3in}\times  \left[-s i \gamma_5\gamma^k A_k \frac{t}{M^2}\left\{
    \frac{i}{4}[\gamma_i,\gamma_j]F_u^{ij}
    \right\}\right]\varphi^\omega_{g,a}(x_4)
   \nonumber\\
   &&\hspace{-3in}= -1_{2\times 2}\{\varphi^\omega_{g,a}(x_4)^\dagger\left[\tau_3\right]\varphi^\omega_{g,a}(x_4)\}   e^{-\frac{t}{M^2}(\omega^2+\bm{p}^2)-t}
   \frac{st}{2M^2} \epsilon_{ijk} A^kF_u^{ij}, 
\end{eqnarray}
and similarly
\begin{eqnarray}
  \varphi^{\rm edge}_{+,e}(x_4)^\dagger \left[-s i \gamma_5\gamma^k A_k e^{-t (1-\frac{(ip^i\gamma_i+D^u)^2+2M\gamma_4\delta(x_4)}{M^2})}\right]
  \varphi^{\rm edge}_{+,e}(x_4)
   \nonumber\\
   &&\hspace{-3in}= -1_{2\times 2}\varphi^{\rm edge}_{+,e}(x_4)^\dagger
   \varphi^{\rm edge}_{+,e}(x_4)  e^{-\frac{t}{M^2}\bm{p}^2}
   \frac{st}{2M^2} \epsilon_{ijk} A^kF_u^{ij}, 
\end{eqnarray}
where $D^u$ and $F_u^{ij}$ are the Dirac operator and field strength for the gauge field $uA$, respectively.
Then we have
\begin{eqnarray}
  {\rm Tr}\left[-s\frac{dH_{DW}(u)}{du}\frac{e^{-t \frac{H_{DW}(u)^2}{M^2}}}{M}\right]
  &=& \int d^4x\;
  \frac{-s}{8\pi^2}
  \frac{\partial}{\partial |x_4|}\left[
    \frac{\sqrt{\pi}}{4\sqrt{t}}e^{-2M|x_4|}\left\{{\rm erfc}\left(\frac{-M|x_4|}{\sqrt{t}}+\sqrt{t}\right)-2\right\}
    \right.\nonumber\\&&\left.+\frac{\sqrt{\pi}}{4\sqrt{t}}e^{2M|x_4|}{\rm erfc}\left(\frac{M|x_4|}{\sqrt{t}}+\sqrt{t}\right)\right]
  \epsilon_{ijk}{\rm tr}_{c}A^kF_u^{ij}, 
\end{eqnarray}
where we have used Eq.~(\ref{eq:spintr}), Eq.~(\ref{eq:pint}), and
\begin{eqnarray}
  \sum_{a=e,o}\sum_{g=\pm}\int_0^\infty d\omega  \varphi^\omega_{g,a}(x_4)^\dagger \left[\tau_3\right] \varphi^\omega_{g',a}(x_4) e^{-\frac{t}{M^2}(\omega^2+M^2)}+\varphi^{\rm edge}_+(x_4)^\dagger\varphi^{\rm edge}_+(x_4)
  \nonumber\\
  = 1_{2\times 2}\frac{1}{4}\frac{\partial}{\partial |x_4|}\left[e^{-2M|x_4|}\left\{{\rm erfc}\left(\frac{-M|x_4|}{\sqrt{t}}+\sqrt{t}\right)-2\right\}
    +e^{2M|x_4|}{\rm erfc}\left(\frac{M|x_4|}{\sqrt{t}}+\sqrt{t}\right)\right].
\end{eqnarray}

Here let us compute
\begin{eqnarray}
        \label{eq:g1}
  g_1(x_4,M) &=&
  \frac{1}{\Gamma\left(\frac{1+s}{2}\right)}\int_0^\infty dt\; t^{\frac{s-1}{2}}
\left[\frac{\sqrt{\pi}}{2\sqrt{t}}e^{2M|x_4|}{\rm erfc}\left(\frac{M|x_4|}{\sqrt{t}}+\sqrt{t}\right)\right]
  \nonumber\\
  &=&
  -\frac{e^{2M|x_4|}}{s\Gamma\left(\frac{1+s}{2}\right)}
  \int_0^\infty dt\; t^{\frac{s}{2}}
  \left[\left(\frac{M|x_4|}{t^{3/2}}-\frac{1}{t^{1/2}}\right)e^{-(M|x_4|/\sqrt{t}+\sqrt{t})^2}\right]
  \nonumber\\
  &=& -\frac{2(M|x_4|)^{(s+1)/2}}{s\Gamma\left(\frac{1+s}{2}\right)}
  \left(K_{(s-1)/2}(2M|x_4|)-K_{(s+1)/2}(2M|x_4|)\right),
\end{eqnarray}
where $K_\nu(z)$ are modified Bessel functions.
$g_1(x_4,M)$ has the following properties.
\begin{eqnarray}
  g_1(0,M) = \frac{1}{s}
  ,\;\;\; g_1(x_4\neq 0,M) =O(1),
\end{eqnarray}
where we have used the expansion
\begin{eqnarray}
K_\nu(x) = x^{-\nu} \left(2^{\nu-1} \Gamma (\nu)+O\left(x^2\right)\right)+x^{\nu} \left(2^{-\nu-1} \Gamma (-\nu)+O\left(x^2\right)\right),
\end{eqnarray}
for small $x$, and
\begin{eqnarray}
 \int_{-\infty}^\infty dx_4 \frac{\partial}{\partial |x_4|}g_1(x_4,M) = -\frac{2}{s}.
\end{eqnarray}
Therefore, we can regard that
\begin{eqnarray}
  \lim_{s\to 0}  s\frac{\partial}{\partial |x_4|}g_1(x_4,M) = -2 \delta(x_4).
\end{eqnarray}
%

Similarly
\begin{eqnarray}
\label{eq:g2}
  g_2(x_4,M) &=& \frac{1}{\Gamma\left(\frac{1+s}{2}\right)}\int_0^\infty dt\; t^{\frac{s-1}{2}}
  \left[\frac{\sqrt{\pi}}{2\sqrt{t}}e^{-2M|x_4|}\left\{{\rm erfc}\left(\frac{-M|x_4|}{\sqrt{t}}+\sqrt{t}\right)-2\right\}\right]
  \nonumber\\
  &=& - \frac{2e^{-2M|x_4|}\sqrt{\pi}}{s\Gamma\left(\frac{1+s}{2}\right)}\lim_{T\to \infty} T^{s/2}
  + \frac{2(M|x_4|)^{\frac{s+1}{2}}}{s\Gamma\left(\frac{1+s}{2}\right)} \left(K_{\frac{s+1}{2}}(2 M|x_4|)+K_{\frac{s-1}{2}}(2 M|x_4|)\right)
  \nonumber\\
\end{eqnarray}
has the following properties,
\begin{eqnarray}
  g_2(0,M) = -\frac{1}{s}
  ,\;\;\; g_2(x_4\neq 0,M) = O(1),
\end{eqnarray}
and 
\begin{eqnarray}
 \int_{-\infty}^\infty dx_4 \frac{\partial}{\partial |x_4|}g_2(x_4,M) = \frac{2}{s}+O(1).
\end{eqnarray}
Therefore, we have
\begin{eqnarray}
  \lim_{s\to 0}  s\frac{\partial}{\partial |x_4|}g_2(x_4,M) = 2\delta(x_4).
\end{eqnarray}

Interestingly, the contribution from $g_1(x_4,M)$ and $g_2(x_4,M)$ cancels.
\begin{eqnarray}
  \lim_{s\to 0}\frac{s}{2}\int_{-\infty}^\infty dx_4 \frac{\partial}{\partial |x_4|}(g_2(x_4,M)+g_1(x_4,M))f(x_4) = 0;
\end{eqnarray}
so does the integrand of the second term in Eq.~(\ref{eq:goalDW}),
\begin{eqnarray}
\label{eq:deta}
\lim_{s\to 0}\frac{1}{\Gamma\left(\frac{1+s}{2}\right)}\int_0^\infty dt\; t^{\frac{s-1}{2}}
{\rm Tr}\left[-s\frac{dH_{DW}(u)}{du}\frac{e^{-t \frac{H_{DW}(u)^2}{M^2}}}{M}\right]
=0,
\end{eqnarray}
at finite $M$.

\subsubsection{Final result of $\eta(H_{DW})$}
Summing up all the contributions, we obtain
\begin{eqnarray}
\label{eq:etaDW}
  \eta(H_{DW}) = \frac{1}{32\pi^2}\int d^4x\;\epsilon(x_4)
  \epsilon_{\mu\nu\rho\sigma}{\rm tr}_{c}F^{\mu\nu}F^{\rho\sigma}
  -\eta(iD^{\rm 3D}),
\end{eqnarray}
where we have added $2[CS|_{x_4=0}/2\pi]$ according to the prescription in Eq.~(\ref{eq:etaandCS}).
The first term of Eq.~(\ref{eq:etaDW}) contains contribution only from the bulk modes;
the edge-localized modes contribute to $g(x_4,M)$, which disappears in the large $M$ limit,
whereas the second term of Eq.~(\ref{eq:etaDW}) entirely comes from the edge-localized modes,
as explicitly computed in the previous subsection.
In the above derivation, Eq.~(\ref{eq:deta}) is particularly important since it is equivalent to
showing
\begin{eqnarray}
 \frac{\partial\eta(H_{DW}(u))}{\partial u} =0.
\end{eqnarray}
Moreover, we can also show
\begin{eqnarray}
 \frac{\partial \eta(H_{DW})}{\partial M} = - \lim_{s\to 0} \frac{s}{M} \frac{1}{32\pi^2}\int d^4x\;\epsilon(x_4)
  \epsilon_{\mu\nu\rho\sigma}{\rm tr}_{c}F^{\mu\nu}F^{\rho\sigma} = 0.
\end{eqnarray}
Namely, our definition of the index is stable against
any variational changes in $M$ and gauge field $A_\mu$.
It only allows discrete jumps by an even integer in
the boundary contribution $\eta(iD^{\rm 3D})$.

\subsection{APS index and physical interpretation}

We have shown that the index is equivalent to that of APS, {\it i.e.}
\begin{eqnarray}
\label{eq:APSfinal}
  \mathcal{I}_{x_4>0} &=& \frac{1}{2}\eta(H_{DW})- \frac{1}{2}\eta(H_{PV})
=  \frac{1}{32\pi^2}\int_{x_4>0} d^4x\; \epsilon_{\mu\nu\rho\sigma}{\rm tr}_cF^{\mu\nu}F^{\rho\sigma}
  -\frac{\eta(iD^{\rm 3D})}{2}.
\end{eqnarray}
If we flip the sign of the Pauli-Villars mass, we obtain the same index in the $x_4<0$ region:
\begin{eqnarray}
  -\mathcal{I}_{x_4<0} &=& \frac{1}{2}\eta(H_{DW})- \frac{1}{2}\eta(H_{PV}|_{M\to -M})
= -\frac{1}{32\pi^2}\int_{x_4<0} d^4x\; \epsilon_{\mu\nu\rho\sigma}{\rm tr}_cF^{\mu\nu}F^{\rho\sigma}
  -\frac{\eta(iD^{\rm 3D})}{2}.\nonumber\\
\end{eqnarray}
In fact, the eta invariant of the domain-wall fermion Dirac operator can be written as the difference 
between the APS indices in the two regions:
\begin{eqnarray}
  \eta(H_{DW}) = \mathcal{I}_{x_4>0} -\mathcal{I}_{x_4<0}.
\end{eqnarray}
Then the sign of the Pauli-Villars mass determines
which is topological and which is normal insulator.
In our computations, we do not need the massless Dirac operator
or global boundary conditions on the fermion fields.
Moreover, we have seen that the eta invariant $\eta(iD^{\rm 3D})$ comes entirely from the
edge-localized modes, while these edge modes do not contribute to the first term of Eq.~(\ref{eq:APSfinal}) at all.

As a final remark of this section, let us consider the second domain-wall
or anti-domain wall, which is needed to compactify our set-up with flat metric.
To define the index, we formally need to consider the domain-wall Dirac fermion operator
in a finite region of $-L<x_4<L$, and, for example, identify the fermion field at $x_4=L$ and $-L$
(periodic boundary condition):
\begin{eqnarray}
      H_{DW} = \gamma_5\left\{D + M\epsilon(x_4)\epsilon(L-x_4)\right\},  
\end{eqnarray}
where the spatial directions are also required to be compactified.
Even in this case, our computation above is valid, at least,
in the large volume limit in the near region of $x_4=0$
and it should be naturally and smoothly continuated to
the another domain-wall at $x_4=L$.
Finally we obtain
\begin{eqnarray}
  \mathcal{I} = \frac{1}{2}\eta(H_{DW})- \frac{1}{2}\eta(H_{PV})
&=&  \frac{1}{32\pi^2}\int_{0<x_4<L} d^4x\; \epsilon_{\mu\nu\rho\sigma}{\rm tr}_cF^{\mu\nu}F^{\rho\sigma}
  \nonumber\\&&-\left.\frac{\eta(iD^{\rm 3D})}{2}\right|_{x_4=0}+\left.\frac{\eta(iD^{\rm 3D})}{2}\right|_{x_4=L}.
\end{eqnarray}

\section{Asymmetric domain-wall}
\label{sec:asymmetricDW}

In the previous section, we have considered
the domain-wall fermion Dirac operator from which
the APS index has been reproduced.
Although the domain-wall fermion is a good model
to describe the topological insulator with boundary,
the size of the fermion gap $|M|$ is the same
both in the normal and topological phases,
which is not generally true in the actual materials.
In this section, we consider a more general case
where the two regions $x_4<0$ and $x_4>0$
have different mass gaps.

\subsection{Effect of asymmetric mass}

Let us consider a modified model with an additional mass $M_2$
without the kink structure,
\begin{eqnarray}
  H_{DW} = \gamma_5(D + M_1\epsilon(x_4)-M_2),
\end{eqnarray}
where both $M_1$ and $M_2$ are positive.
This introduces the asymmetric mass to the fermion in
normal and topological phases,
and a step-function-like term in $H_{DW}^2$ (when $\partial_{x_4}B=0$),
\begin{eqnarray}
  H_{DW}^2 = -\partial_4^2 + B^2 + M_1^2+M_2^2 - 2M_1\gamma_4\delta(x_4)-2M_1M_2\epsilon(x_4).
\end{eqnarray}

Because of the step function, there are three types of eigensolutions of $H_W^2$
using the same decomposition $\varphi_\pm (x_4)\otimes\phi_\lambda^{\rm 3D}(\vec{x})$, where
$iD^{\rm 3D}\phi_\lambda^{\rm 3D}(\vec{x})=\lambda\phi_\lambda^{\rm 3D}(\vec{x})$ and
$\tau_3\varphi_\pm (x_4)=\pm \varphi_\pm (x_4)$
as in the previous section;
1) localized bound state (edge state),
\begin{eqnarray}
  \varphi^{\rm edge}_{+} (x_4) &=& \left\{
\begin{array}{cc}
  u_+\sqrt{\frac{M_1^2-M_2^2}{M_1}}e^{-(M_1-M_2)x_4} & (x_4\geq 0)\\
  u_+\sqrt{\frac{M_1^2-M_2^2}{M_1}}e^{(M_1+M_2)x_4} & (x_4<0)
\end{array}\right.,
\end{eqnarray}
where the eigenvalue of $H_W^2$ is $\Lambda^2=\lambda^2$,
2) plane waves extended only in the $x_4>0$ region,
\begin{eqnarray}
  \varphi^{\omega}_{\pm} (x_4) &=& \left\{
\begin{array}{cc}
  \frac{u_\pm}{\sqrt{2\pi(\omega^2+\mu_{\pm}^2)}}
  \left\{(i\omega+\mu_\pm)e^{i\omega x_4}+(i\omega- \mu_\pm)e^{-i\omega x_4}\right\}, & (x_4\geq 0)\\
  u_\pm\frac{2i\omega}{\sqrt{2\pi(\omega^2+\mu_{\pm}^2)}}e^{\Omega x_4} & (x_4<0)
\end{array}\right.,
\end{eqnarray}
where $\omega=\sqrt{\Lambda^2-\lambda^2-(M_1-M_2)^2}$,
$\Omega=\sqrt{-\Lambda^2+\lambda^2+(M_1+M_2)^2}$,
and $\mu_\pm=\Omega\mp 2M_1$,
and 3) plane waves extended in the whole region.
\begin{eqnarray}
  \label{eq:sol3}
  \varphi^{\omega}_{\pm} (x_4) &=& \left\{
\begin{array}{cc}
  u_\pm(A e^{i\omega_1 x_4}+B e^{-i\omega_1 x_4}) & (x_4\geq 0)\\
  u_\pm(C e^{i\omega_2 x_4}+D e^{-i\omega_2 x_4}) & (x_4<0)
\end{array}\right.,
\end{eqnarray}
where $\omega_1=\sqrt{\Lambda^2-\lambda^2-(M_1-M_2)^2}$,
$\omega_2=\sqrt{\Lambda^2-\lambda^2-(M_1+M_2)^2}$,
and the coefficients satisfy $A+B=C+D$, and $-i\omega_1(A-B)+i\omega_2(C-D)\mp 2M_1(A+B)=0$.
The orthonormality of the above eigenfunctions can be confirmed using the relation \cite{Trott}\footnote{
We thank H.~Nakazato and M.~Ochiai for useful information about the system in a step potential.
}
\begin{eqnarray}
  \int_0^\infty dx_4 e^{i\omega x_4} = \pi\delta(\omega) +i \mathcal{P}\frac{1}{\omega},
\end{eqnarray}
where $\mathcal{P}$ denotes the principal value.

It is important to note that the above solutions all satisfy
the nontrivial boundary condition
\begin{eqnarray}
  -\lim_{\epsilon \to 0}(\partial_{x_4}\varphi^{\omega/{\rm edge}}_\pm(+\epsilon)
  -\partial_{x_4}\varphi^{\omega/{\rm edge}}_\pm(-\epsilon))
  \mp 2M_1\varphi^{\omega/{\rm edge}}_\pm(0) = 0,
\end{eqnarray}
which respects the $SO(3)$ rotational symmetry on the surface.
It is also important to note that the edge mode exists only when $M_1>M_2$,
otherwise the above solution is not normalizable.

An appropriate Pauli-Villars operator
in the case of $M_1>M_2$ is given by
\begin{eqnarray}
  H_{PV} = \gamma_5(D - M_1 + M_2\epsilon(x_4)),
\end{eqnarray}
whose total mass $- M_1 + M_2\epsilon(x_4)$ does not change
its sign at $x_4$ and hence does not develop any edge-localized modes.

Since the additional mass $M_2$ does not break the $\gamma_5$ Hermiticity
of the domain-wall and Pauli-Villars Dirac operators,
we can define the index
\begin{eqnarray}
  \mathcal{I} &=& \frac{1}{2}\eta(H_{DW})- \frac{1}{2}\eta(H_{PV}).
\end{eqnarray}
Furthermore, we can show 
\begin{eqnarray}
  \frac{d\mathcal{I}}{dM_2}
  = \lim_{s\to 0} s \times (\mbox{finite terms}) = 0,
\end{eqnarray}
for $M_1>M_2$, from which there is no doubt that $\mathcal{I}$ is equivalent to the APS index.
It is still instructive to directly compute the index in an extreme case,
where the mass gap in the $x_4<0$ region is infinitely large,
and all the wave functions are constrained to the $x_4\geq 0$ region.
This is equivalent to considering
the original system of manifold with boundary, as a closed system
(similar studies were done in Refs.~\cite{Marachevsky:2003zb, Beneventano:2004yy}).

\subsection{Shamir-type domain-wall}

In the following, let us take an extreme limit
where $M_1+M_2 =\infty$, while $M_1-M_2=M$ is fixed.
In this case, we can safely neglect the type 3) plane wave solutions 
in Eq.~(\ref{eq:sol3}) and the other two types of eigenfunctions become
\begin{eqnarray}
  \varphi^{\rm edge}_{+} (x_4) &=& \left\{
\begin{array}{cc}
  u_+\sqrt{2M}e^{-M x_4} & (x_4\geq 0)\\
  0 & (x_4<0)
\end{array}\right.,
\end{eqnarray}
where $\Lambda^2=\lambda^2$ is unchanged, and 
\begin{eqnarray}
  \varphi^{\omega}_+ (x_4) &=& \left\{
\begin{array}{cc}
  \frac{u_+}{\sqrt{2\pi(\omega^2+M^2)}}
  \left\{(i\omega-M)e^{i\omega x_4}+(i\omega+M)e^{-i\omega x_4}\right\} & (x_4\geq 0)\\
  0 & (x_4<0)
\end{array}\right.,
\end{eqnarray}
\begin{eqnarray}
  \varphi^{\omega}_- (x_4) &=& \left\{
\begin{array}{cc}
  \frac{u_-}{\sqrt{2\pi}}
  \left(e^{i\omega x_4}-e^{-i\omega x_4}\right) & (x_4\geq 0)\\
  0 & (x_4<0)
\end{array}\right.,
\end{eqnarray}
where $\omega=\sqrt{\Lambda^2-\lambda^2-M^2}$.
The above wave functions are equivalent to the complete set of
the massive Dirac operator,
\begin{eqnarray}
H_+ = \gamma_5(D+M),
\end{eqnarray}
extending only in the $x_4 \geq 0$ region, whose boundary condition is
locally given by
\begin{eqnarray}
  \varphi_-|_{x_4=0}=0,\;\;\; (\partial_{x_4}+M)\varphi_+|_{x_4=0}=0.
\end{eqnarray}
In fact, this system corresponds to the so-called Shamir-type domain-wall fermion
\cite{Shamir:1993zy,Furman:1994ky}.

In the same way, the complete set of the  Pauli-Villars operator
converges to that of 
\begin{eqnarray}
H_- = \gamma_5(D-M),
\end{eqnarray}
which are given by
\begin{eqnarray}
  \varphi^{\omega}_{PV +} (x_4) &=& 
  \frac{u_+}{\sqrt{2\pi(\omega^2+M^2)}}
  \left\{(i\omega+M)e^{i\omega x_4}+(i\omega-M)e^{-i\omega x_4}\right\}\;\;\; (x_4\geq 0)\\
  \varphi^{\omega}_{PV -} (x_4) &=& 
  \frac{u_-}{\sqrt{2\pi}}
  \left(e^{i\omega x_4}-e^{-i\omega x_4}\right) \;\;\; (x_4\geq 0),
\end{eqnarray}
satisfying another local  boundary condition
\begin{eqnarray}
  \varphi_{PV-}|_{x_4=0}=0,\;\;\; (\partial_{x_4}-M)\varphi_{PV+}|_{x_4=0}=0.
\end{eqnarray}

Now we can explicitly compute with these complete sets the index
\begin{eqnarray}
  \label{eq:indexShamirDW}
  \mathcal{I} &=& \frac{1}{2}\eta(H_+)- \frac{1}{2}\eta(H_-).
\end{eqnarray}
In fact, except that
the variation of the eta invariants does not vanish separately,
\begin{eqnarray}
 \int_0^1 du \frac{d}{du}\eta(H_\pm) &=& \int_0^1 du\lim_{s\to 0}\frac{1}{\Gamma\left(\frac{1+s}{2}\right)}\int_0^\infty dt\; t^{\frac{s-1}{2}}
       {\rm Tr}\left[\mp s\frac{dH_\pm(u)}{du}\frac{e^{-t \frac{H_\pm(u)^2}{M^2}}}{M}\right]
       \nonumber\\  &=&
       \mp \lim_{s\to 0} \int_0^1 du\int_{x_4>0} d^4x\;
  \frac{s}{8\pi^2}
  \frac{\partial}{\partial x_4}g_\pm(x_4,M)
  \epsilon_{ijk}{\rm tr}_{c}A^kF_u^{ij}
  \nonumber\\  &=&
  \mp \lim_{s\to 0}\int_0^1 du \lim_{\epsilon\to 0}\int_{x_4>0} d^4x\;
  \frac{s}{8\pi^2}
  \left(\pm \frac{\delta(x_4-\epsilon)}{s}\right)
  \epsilon_{ijk}{\rm tr}_{c}A^kF_u^{ij}
  \nonumber\\  &=&
  - \left.\frac{CS}{2\pi}\right|_{x_4=0},
\end{eqnarray}
where $g_+(x_4,M)=g_1(x_4,M)$, and $g_-(x_4,M)=g_2(x_4,M)$, already
appeared in Eqs.~(\ref{eq:g1}) and (\ref{eq:g2}), respectively,
the computation is very similar to the one
obtained in the previous section.
The results are summarized as
\begin{eqnarray}
  \eta(H_+) &=& \frac{1}{32\pi^2}\int_{x_4>0} d^4x\;
  \epsilon_{\mu\nu\rho\sigma}{\rm tr}_{c}F^{\mu\nu}F^{\rho\sigma}
  -\frac{1}{\pi}CS|_{x_4=0}+2\left[\frac{1}{2\pi}CS|_{x_4=0}\right],\\
  \eta(H_-) &=& -\frac{1}{32\pi^2}\int_{x_4>0} d^4x\;
  \epsilon_{\mu\nu\rho\sigma}{\rm tr}_{c}F^{\mu\nu}F^{\rho\sigma},
\end{eqnarray}
where we have again put the Gauss symbol term $2\left[CS|_{x_4=0}/2\pi \right]$
to maintain the gauge invariance.
Thus, $\mathcal{I}$ turns out to be the same index as the original
APS in Eq.~(\ref{eq:indexShamirDW}).

\section{Summary and discussion}
\label{sec:summary}
In this work, we have tried to describe the APS index theorem
in a ``physicist-friendly'' way in a simple set-up with a flat metric,
for the Dirac fermion operator with $U(1)$ or $SU(N)$ gauge field background.
Our method corresponds to a generalization of the Fujikawa method on closed
manifolds to that on manifolds with boundaries.

First, we have revisited the original set-up by APS and reproduced
the index theorem in an adiabatic expansion.
Contrary to the intuition that the eta invariant is a contribution
of the edge-localized modes, we have found that
the APS boundary condition allows no such edge modes to exist.
Instead, a non-trivial pole structure of the coefficients
of the bulk extended modes produces it.
We have also discussed that the APS boundary is unnatural
and unlikely to be realized in actual materials with boundary.

Then we have discussed what is required in more physical set-ups.
In physics, what we call boundary is actually a domain-wall on which
some physical parameter becomes discontinuous.
Every topological insulator is non-trivial only when
it is surrounded by normal insulators.
It is more natural to consider a massive fermion since it is not
the helicity but rotational symmetry that should be preserved on a surface.
Any boundary condition should not be imposed by hand but
should be given by the local dynamics of the system.
We have concluded that the domain-wall Dirac fermion operator is 
a good candidate to reformulate the index theorem in physics.

Next, we have defined a new index by the eta invariant of
the four-dimensional domain-wall fermion Dirac operator with its Pauli-Villars regulator.
The kink structure in the mass term automatically forces the
fermion fields to satisfy a boundary condition, which is locally given
and respects the $SO(3)$ rotational symmetry on the surface.
As a consequence, the edge-localized modes appear in the complete set of the free Dirac operator.
We have applied the Fujikawa method to this complete set satisfying
the non-trivial boundary condition.
Since the boundary condition is no longer dependent of gauge fields,
we do not need the  adiabatic approximation.
We have obtained an index, which
is stable against the changes of mass and gauge field.
This new index coincides with the APS index.
Moreover, in our set-up, the physical origin of the eta invariant is clearer.
It comes entirely from the edge-localized modes.

Finally, we have considered the case with asymmetric masses
and computed the index in the limit where one of the masses goes to infinity.
This case is closer to the original set-up by APS,
where we do not need to consider the $x_4<0$ region.
In lattice gauge theory, this extremal case is known as
the Shamir-type domain-wall fermion.
We have confirmed by the direct computation that the index remains
the same as the original APS index.

In this work, we have employed the Pauli-Villars regularization.
It is interesting to give
a non-perturbative definition of the APS index
based on the lattice regularization 
as was done for the AS index \cite{Hasenfratz:1998ri}.
As the Wilson fermion Dirac operator has the $\gamma_5$ Hermiticity
and its determinant is real, we would be able to define an index by
$\eta(\gamma_5(D_W+M\epsilon(x_4)))/2$
(assuming that the Wilson term has the opposite sign to the mass $M$),
which coincides with the APS index, at least, in the continuum limit.
In the lattice regularization, one would be able to
increase the effective number of flavors $N_f$, by tuning the mass and Wilson term,
so that some of the doubler modes become physical \cite{Golterman:1992ub}. 
Even in that case, the APS index would be unchanged except for
the overall multiplication of $N_f$.

The APS index theorem describes a part of the anomaly descent equations
\cite{Stora:1983ct,Zumino:1983ew,Zumino:1983rz, AlvarezGaume:1983cs, Sumitani:1984ed}, 
in which the parity anomaly or the CS term in $2n+1$ dimensions
appears as the surface term of the axial $U(1)$ anomaly in $2n+2$ dimensions.
Our work describing the same index in terms of the domain-wall Dirac operator
corresponds to its fermionic expression.
It is interesting to extend our work to the $2n$-dimensional
Weyl fermion system, which appears as the
edge-localized state of the $2n+1$-dimensional gapped bulk fermions.
As already investigated in the literature
\cite{AlvarezGaume:1985di, DellaPietra:1986qd, Kaplan:1995pe}, 
the gauge anomaly should be canceled by the surface contribution from the bulk $\eta$-invariant.

A further interesting question is whether we can incorporate the full set of
anomaly descent equations in the $2n+2\to 2n+1\to 2n$ dimensions,
in one Dirac fermion operator.
In the conventional approach with the manifold with boundary,
this is impossible since the boundary of the boundary must be trivial as a consequence of the homology.
Combining two domain-walls having different quantum numbers,
however, we have already proposed such an interesting ``doubly gapped''
fermion system \cite{Fukaya:2016ofi},
where the edge-of-edge state \cite{Hashimoto:2017tuh}
appears only at the junction of the domain-walls.
Our one-loop level computation shows that
the structure of the full set of anomaly descent equations is embedded in the fermion determinant.
The current work would provide a mathematical basis for investigating
such non-trivial domain-wall systems.


\acknowledgements
We thank K. Hashimoto for organizing the study group on topological
insulators and useful discussions.
We also thank S.~Aoki, M.~L\"uscher, T.~Misumi, H.~Suzuki and A.~Tanaka for discussions.
This work is supported in part by the Japanese Grant-in-Aid for
Scientific Research(Grants No. JP26247043 and No. JP15K05054).

\appendix

\section{Example in two dimensions}
\label{app:2dexample}

\newcommand{\DI}{D^{\mathrm{1D}}}
\newcommand{\DII}{D^{\mathrm{2D}}}
\newcommand{\del}{\partial}
\newcommand{\Zb}{\mathbb{Z}}
\newcommand{\sign}{\mathrm{sign}}
\newcommand{\ind}{\mathrm{ind}}
\renewcommand{\Re}{\mathrm{Re\,}}

\begin{figure}
  \centering
  \includegraphics[width=9cm]{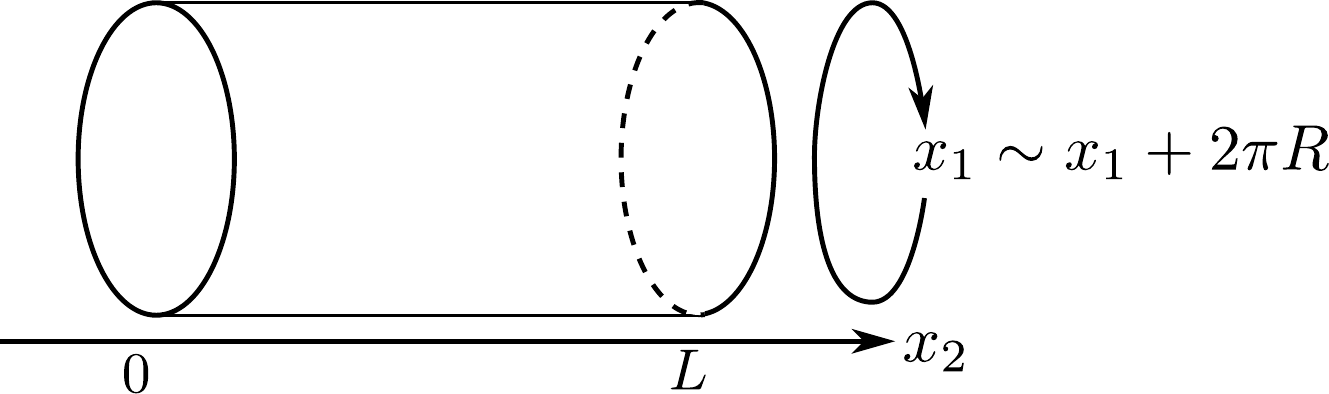}
  \caption{two-dimensional cylinder.}
  \label{fig:cylinder}
\end{figure}
In this appendix, we consider an example of $U(1)$ gauge theory
in two dimensions with boundary.
Under a constant background magnetic field,
we non-perturbatively confirm the APS index theorem, discussed in
Sec. \ref{sec:massless}.

Let us consider a two-dimensional cylinder parameterized by $(x_1,x_2)$
as depicted in Fig.~\ref{fig:cylinder}.
Here $x_1,\ (x_1\sim x_1+2\pi R)$ parametrizes the circle of radius $R$,
while $x_2,\ (0\le x_2 \le L)$ parametrizes a segment of length $L$.
This cylinder has two disconnected circular boundaries at $x_2=0$ and $x_2=L$.

We introduce a constant magnetic field $F_{12}=B$ on this cylinder.
We choose the Landau gauge and the vector potential is written as
\begin{align}
  A_1=-Bx_2+\frac{a_0}{R},\quad
  A_2=0,\label{vectorpotential2d}
\end{align}
where $a_0$ is a constant.
This constant $a_0$ is the holonomy around the circle at $x_2=0$ boundary,
which corresponds to the Chern-Simons term in one dimension:
\begin{align}
  \frac{1}{2\pi}\int dx_1 A_1(x_1,x_2=0) 
  =a_0\;\;\; \left(=\frac{CS}{2\pi}\right).
\end{align}
Similarly we define the holonomy $a_L$ at $x_2=L$ by
\begin{align}
  \frac{1}{2\pi}\int dx_1 A_1(x_1,x_2=L)
  =:a_L.
\end{align}
A useful relation obtained from  Eq.~\eqref{vectorpotential2d}
using the Stoke's theorem is
\begin{align}
  a_L
  =-\frac{1}{2\pi}\int d^2x F_{12}+a_0.
  \label{aLa0relation}
\end{align}

We consider the Dirac operator on this cylinder given by
\begin{align}
  \DII=\gamma_1 D_1+\gamma_2 D_2,
\end{align}
where $\gamma_i,\ (i=1,2)$ are $2\times 2$ gamma matrices of two dimensions
which satisfy $\{\gamma_i,\gamma_j\}=2\delta_{ij}$,
and $D_i=\del_i+iA_i,\ (i=1,2)$ are covariant derivatives.
We also introduce the chirality matrix $\gamma_3$ by $\gamma_3=i\gamma_1\gamma_2$. 

\subsection{APS index theorem in the two-dimensional example}

We count the index
\begin{align}
  \ind(\DII):=n_{+}-n_{-},\qquad
  n_{\pm}:=(\text{number of zero-modes with } \gamma_3=\pm1),
\end{align}
by explicitly constructing the zero-mode wave functions under the APS boundary condition
and confirm that the APS index theorem holds.

Since we have translation symmetry in the $x_1$ direction
we can write the zero-mode wave functions as
\begin{align}
  \psi_{n\pm}(x_1,x_2)=e^{i\frac{n}{R}x_1}
  \phi_{n\pm}(x_2),\quad n\in\Zb,
\end{align}
where $\pm$ stands for the chirality, i.e. $\gamma_3\psi_{n\pm}=\pm \psi_{n\pm}$ and
 $\gamma_3\phi_{n\pm}=\pm \phi_{n\pm}$.  Then the zero-mode equation $\DII \psi_{n\pm}=0$
implies
\begin{align}
  \phi_{n\pm}(x_2)=\exp\left[
  \pm\frac{1}{2B}\left(
  Bx_2-\frac{n+a_0}{R}
  \right)^2
  \right]\chi_{n\pm},
  \label{zeromodewavefunction}
\end{align}
where $\chi_{n\pm}$ is a constant spinor which satisfies $\gamma_3\chi_{n\pm}=\pm\chi_{n\pm}$.
Thus once $n$ and the chirality $\pm$ is given, the zero-mode wave function, if ever exists,
is fixed up to overall constant.

Let us next consider the boundary condition
at $x_2=0$.
This boundary condition is $\psi(x_2=0)=0$
if eigenvalue of $i\gamma_3 D_1|_{x_2=0}$ is positive. 
Notice that if a zero-mode eigenfunction $\psi_{n\pm}(x)$ satisfy $\psi_{n\pm}(x_2=0)=0$
then $\psi_{n\pm}(x)=0$ for all $x_2$ as seen from Eq.~\eqref{zeromodewavefunction}.
Therefore the zero-modes which survive after imposing the APS boundary condition at $x_2=0$ are
\begin{align}
  \psi_{n+},\ (n+a_0>0),\qquad \psi_{n-},\ (n+a_0<0). \label{APSat0}  
\end{align}

Let us turn to the APS boundary condition at $x_2=L$.
Since the orientation is opposite, the APS boundary condition is $\psi(x_2=L)=0$
if the eigenvalue of $-i\gamma_3 D_1|_{x_2=L}$ is positive.
The surviving zero-modes are
\begin{align}
  \psi_{n+},\ (n+a_L<0),\qquad \psi_{n-},\ (n+a_L>0). \label{APSatL} 
\end{align}

Finally let us combine both conditions eqs.~\eqref{APSat0},\eqref{APSatL}.
The surviving zero-modes are
\begin{align}
  \psi_{n+},\ (-a_0<n<-a_L),\qquad \psi_{n-},\ (-a_L<n<-a_0). \label{wavefunctioncondition} 
\end{align}
As a result the number of zero-modes $n_{\pm}$ is given by
\begin{align}
  n_{+}&=(\text{number of integers } n,\quad -a_0<n<-a_L),\nonumber\\
  n_{-}&=(\text{number of integers } n,\quad -a_L<n<-a_0).\label{APS2D}
\end{align}

When $B>0$ the inequality $-a_0<-a_L$ holds from Eq.~\eqref{aLa0relation}
and the numbers of zero-modes $n_{\pm}$ given in Eq.~\eqref{APS2D} read
\begin{align}
  n_{+}=[a_0]-[a_L],\quad n_{-}=0, 
\end{align}
where $[\cdot]$ is the Gauss symbol.
The index is rewritten by making use of Eq.~\eqref{aLa0relation} as
\begin{align}
  \ind(\DII)=n_{+}-n_{-}
  =[a_0]-[a_L]=-a_0+a_L+\frac{1}{2\pi}\int d^2x F_{12}+[a_0]-[a_L].  
  \label{ind2D2}
\end{align}
As we will see in Eqs.~\eqref{result1Deta1} and \eqref{result1Deta2},
the one-dimensional eta invariants are written in terms of $a_0,a_L$ as
\begin{align}
  \eta(iD_1|_{x_2=0},0)=-1+2(a_0-[a_0]),\quad  
  \eta(iD_1|_{x_2=L},0)=-1+2(a_L-[a_L]).
  \label{eta1Datboundary}
\end{align}
Substituting Eq.~\eqref{eta1Datboundary} into Eq.~\eqref{ind2D2}, we obtain
\begin{align}
  \ind(\DII)=\frac{1}{2\pi}\int d^2x F_{12}-\frac12\Big(\eta(iD_1|_{x_2=0},0)-\eta(iD_1|_{x_2=L},0)\Big). \label{indextheorem2D}
\end{align} 
This is nothing but the APS index theorem.
Note that the result is unchanged even for $B<0$.

Let us mention an interesting observation about the zero-mode wave function.
The condition \eqref{wavefunctioncondition} derived from APS boundary condition
is equivalent to requiring that the wave function \eqref{zeromodewavefunction} is Gaussian
and the peak of this Gaussian wave function sits between two boundaries.
This observation may be a hint to find
a physical interpretation of the APS boundary condition.

\subsection{Eta invariant in 1 dimension}
We consider one-dimensional circle parametrized by $x_1\sim x_1+2\pi R$
and the $U(1)$ gauge field on it.
The one-dimensional Dirac operator is a simple covariant derivative, which is written as
\begin{align}
  \DI=\del_{1}+iA_{1}.
\end{align}
We choose the gauge in which $A_{1}$ is a constant,
whose integral gives a non-trivial Chern-Simons term in one dimension:
\begin{align}
  A_1=\frac{a}{R}\ \Rightarrow\ \oint A_{1}dy=2\pi R A_{1}=2\pi a (= CS).
\end{align}
The eigenvalues of $-i\DI$ are
\begin{align}
  \lambda_{n}=\frac{n+a}{R},\qquad n\in \Zb,
\end{align}
where the $n$-th eigen-function is given by $e^{inx_1}$.
The eta invariant (with finite $s$) is defined as
\begin{align}
\eta(-i\DI,s)=\sum_{n\in \Zb}\sign(\lambda_n)
\frac{1}{|\lambda_n|^s}
\label{1dim-eta},
\end{align}
and we take the $s\to 0$ limit.

\subsubsection{Evaluation by Hurwitz zeta function}
If $a$ is not an integer, then by a large gauge transformation $a$ can be chosen such that
\begin{align}
  0<a<1.
\end{align}
In this gauge choice the eta invariant \eqref{1dim-eta} becomes
\begin{align}
  \eta(-i\DI,s)=\sum_{n=0}^{\infty}\frac{1}{(n+a)^s}-\sum_{n=-1}^{-\infty}\frac{1}{(-(n+a))^s}
  =\zeta(s,a)-\zeta(s,1-a),\label{etazeta}
\end{align}
where $\zeta(s,a)$ is the Hurwitz zeta function given for $\Re s >1 $ by
\begin{align}
  \zeta(s,a)=\sum_{n=0}^{\infty}\frac{1}{(n+a)^s}.
\end{align}
$\zeta(s,a)$ for $\Re s \le 1$ is defined by the analytic continuation from $\Re s >1 $.

It is known that for $a>0$
(see for example, Ch.12 of \cite{Apostol})
\begin{align}
  \zeta(0,a)=\frac12 -a.
\end{align}
Since we choose the gauge in which $a>0,\ 1-a >0$
we can apply this equation to \eqref{etazeta} and obtain
\begin{align}
  \eta(-i\DI,0)=1-2a.
\end{align}
For a generic gauge, the eta invariant is written as
\begin{align}
  \eta(-i\DI,0)=1-2(a-[a]), \quad a=\frac{1}{2\pi}\oint A,
  \label{result1Deta1}
\end{align}
where $[\cdot]$ is the Gauss symbol.
Notice that $\eta(-i\DI,0)$ is gauge invariant as we expected,
although $a$ is not gauge invariant.
In the prescription  by APS, the zero eigenvalue is considered to be ``positive''.
Therefore, the $a\to 0$ limit should be taken from the positive side,
leading to $\eta(-i\DI,0)|_{a\to 0}\to 1$.

\subsubsection{Manifestly gauge invariant calculation}
Here we show more explicit and manifestly gauge invariant calculation of the eta invariant.
We may rewrite $\eta(-i\DI,s)$ in the following way.
\begin{align}
  \eta(-i\DI,s)&=\sum_{n\in\Zb} \frac{\lambda_n}{|\lambda_n|^{1+s}}\nonumber\\
  &=\frac{1}{\Gamma\left(\frac{s+1}{2}\right)}\int_0^{\infty}dt \,
  t^{\frac{s-1}{2}}
  \sum_{n\in\Zb}\lambda_n
  e^{-t\lambda_n^2}\nonumber\\
  &=\frac{1}{\Gamma\left(\frac{s+1}{2}\right)}\int_0^{\infty}dt \,
  t^{\frac{s-1}{2}}
  \sum_{n\in\Zb}\frac{(n+a)}{R}
  e^{-t(\frac{n+a}{R})^2}.
  \label{etacalc1}
\end{align}
Let $\tilde{g}(k)$ be
\begin{align}
  \tilde{g}(k):=\frac{(k+a)}{R}e^{-t(\frac{k+a}{R})^2},
\end{align}
and apply the Poisson resummation formula
\begin{align}
  \sum_{n\in\Zb} \tilde{g}(n)
  =\sum_{n\in\Zb} 2\pi g(2\pi n).
\end{align}
Here $g(x)$ is the Fourier transformation of $\tilde{g}(k)$.
This $g(x)$ is calculated as
\begin{align}
  2\pi g(x)=\int dk e^{ikx}\tilde{g}(k)
  =\frac{R^2\sqrt{\pi}}{2t^{3/2}} ix e^{-iax}e^{-\frac{R^2x^2}{4t}}.
\end{align}
Thus
\begin{align}
  \sum_{n\in\Zb}\tilde{g}(n)=\sum_{n\in\Zb}2\pi g(2\pi n)
  =\sum_{n\in\Zb} \frac{R^2\sqrt{\pi}}{2t^{3/2}}
   2\pi in e^{-2\pi ina}
  e^{-\frac{4\pi^2 R^2 n^2}{4t}}.
\end{align}
Substituting this relation to Eq.~\eqref{etacalc1}, we obtain
\begin{align}
  \eta(-i\DI,s)&=\frac{1}{\Gamma\left(\frac{s+1}{2}\right)}\int_0^{\infty}dt \,
  t^{\frac{s-1}{2}}\sum_{n\in\Zb}
  \frac{R^2\sqrt{\pi}}{2t^{3/2}}
  2\pi i n e^{-2\pi ina}
  e^{-\frac{\pi^2 R^2 n^2}{t}}\nonumber\\
  &=\frac{R^2 2\pi \sqrt{\pi}}{\Gamma\left(\frac{s+1}{2}\right)}
  \sum_{n=1}^{\infty} n \sin(2\pi a n) A_{n},
\end{align}
where $A_n$ is defined and calculated as
\begin{align}
  A_n=\int_0^{\infty}dt t^{\frac{s}{2}-2}e^{-\frac{\pi^2 R^2 n^2}{t}}
  =\Gamma\left(-\frac{s}{2}+1\right)(\pi^2 R^2 n^2)^{\frac{s}{2}-1}.
\end{align}
Then $\eta(-i\DI,0)$ is calculated as
\begin{align}
\eta(-i\DI,0)
&=\frac{2}{\pi}\sum_{n=1}^{\infty}\frac{\sin(2\pi a n)}{n}
  =
1-2(a-[a]),
  \label{result1Deta2}
\end{align}
where the last equality is obtained from the Fourier transformation of
the linear function in a range $0<a<1$ and extend it to the whole region,
using the periodicity in $a\to a+$ integers.
The limit where $a$ goes to an integer should be taken from the positive side. 
This result is the same as Eq.~\eqref{result1Deta1} as expected.



\begin{thebibliography}{99}

\bibitem{Atiyah:1963zz}
  M.~F.~Atiyah and I.~M.~Singer,
  ``The index of elliptic operators on compact manifolds,''
  Bull.\ Am.\ Math.\ Soc.\  {\bf 69}, 422 (1963).
  doi:10.1090/S0002-9904-1963-10957-X

\bibitem{Atiyah:1968mp}
  M.~F.~Atiyah and I.~M.~Singer,
  ``The Index of elliptic operators. 1,''
  Annals Math.\  {\bf 87}, 484 (1968).
  doi:10.2307/1970715

\bibitem{Jackiw:1977pu}
  R.~Jackiw and C.~Rebbi,
  ``Spinor Analysis of Yang-Mills Theory,''
  Phys.\ Rev.\ D {\bf 16}, 1052 (1977).
  doi:10.1103/PhysRevD.16.1052

\bibitem{Fujikawa:1979ay}
  K.~Fujikawa,
  ``Path Integral Measure for Gauge Invariant Fermion Theories,''
  Phys.\ Rev.\ Lett.\  {\bf 42}, 1195 (1979).
  doi:10.1103/PhysRevLett.42.1195


\bibitem{Atiyah:1975jf}
  M.~F.~Atiyah, V.~K.~Patodi and I.~M.~Singer,
  ``Spectral asymmetry and Riemannian Geometry I,''
  Math.\ Proc.\ Cambridge Phil.\ Soc.\  {\bf 77}, 43 (1975).
  doi:10.1017/S0305004100049410

\bibitem{Atiyah:1976jg} 
  M.~F.~Atiyah, V.~K.~Patodi and I.~M.~Singer,
  ``Spectral asymmetry and Riemannian geometry II,''
  Math.\ Proc.\ Cambridge Phil.\ Soc.\  {\bf 78}, 405 (1975).
  doi:10.1017/S0305004100051872

\bibitem{Atiyah:1980jh} 
  M.~F.~Atiyah, V.~K.~Patodi and I.~M.~Singer,
  ``Spectral asymmetry and Riemannian geometry. III,''
  Math.\ Proc.\ Cambridge Phil.\ Soc.\  {\bf 79}, 71 (1976).
  doi:10.1017/S0305004100052105
  
\bibitem{Stora:1983ct}
  R.~Stora,
  ``Algebraic Structure And Topological Origin Of Anomalies,''
  Progress in Gauge Field Theory. NATO ASI Series (Series B: Physics), vol 115. Springer, Boston, MA. (1984)
  doi:10.1007/978-1-4757-0280-4\_19

\bibitem{Zumino:1983ew}
  B.~Zumino,
  ``Chiral Anomalies And Differential Geometry: Lectures Given At Les Houches, August 1983,''
  In *Treiman, S.b. ( Ed.) Et Al.: Current Algebra and Anomalies*, 361-391 and Lawrence Berkeley Lab. - LBL-16747 (83,REC.OCT.) 46p (1985)

\bibitem{Zumino:1983rz}
  B.~Zumino, Y.~S.~Wu and A.~Zee,
  ``Chiral Anomalies, Higher Dimensions, and Differential Geometry,''
  Nucl.\ Phys.\ B {\bf 239}, 477 (1984).
  doi:10.1016/0550-3213(84)90259-1

\bibitem{AlvarezGaume:1983cs} 
  L.~Alvarez-Gaume and P.~H.~Ginsparg,
  ``The Topological Meaning of Nonabelian Anomalies,''
  Nucl.\ Phys.\ B {\bf 243}, 449 (1984).
  doi:10.1016/0550-3213(84)90487-5

\bibitem{Sumitani:1984ed} 
  T.~Sumitani,
  ``Chiral Anomalies and the Generalized Index Theorem,''
  J.\ Phys.\ A {\bf 17}, L811 (1984).
  doi:10.1088/0305-4470/17/14/016

  

\bibitem{Redlich:1983dv}
  A.~N.~Redlich,
  ``Parity Violation and Gauge Noninvariance of the Effective Gauge Field Action in Three-Dimensions,''
  Phys.\ Rev.\ D {\bf 29}, 2366 (1984).
  doi:10.1103/PhysRevD.29.2366

\bibitem{Niemi:1983rq}
  A.~J.~Niemi and G.~W.~Semenoff,
  ``Axial Anomaly Induced Fermion Fractionization and Effective Gauge Theory Actions in Odd Dimensional Space-Times,''
  Phys.\ Rev.\ Lett.\  {\bf 51}, 2077 (1983).
  doi:10.1103/PhysRevLett.51.2077

\bibitem{AlvarezGaume:1984nf}
  L.~Alvarez-Gaume, S.~Della Pietra and G.~W.~Moore,
  ``Anomalies and Odd Dimensions,''
  Annals Phys.\  {\bf 163}, 288 (1985).
  doi:10.1016/0003-4916(85)90383-5



\bibitem{Gromov:2015fda}
  A.~Gromov, K.~Jensen and A.~G.~Abanov,
  ``Boundary effective action for quantum Hall states,''
  Phys.\ Rev.\ Lett.\  {\bf 116}, no. 12, 126802 (2016)
  doi:10.1103/PhysRevLett.116.126802
  [arXiv:1506.07171 [cond-mat.str-el]].

\bibitem{Witten:2015aba}
  E.~Witten,
  ``Fermion Path Integrals And Topological Phases,''
  Rev.\ Mod.\ Phys.\  {\bf 88}, no. 3, 035001 (2016)
  doi:10.1103/RevModPhys.88.035001
  [arXiv:1508.04715 [cond-mat.mes-hall]].

\bibitem{Metlitski:2015yqa}
  M.~A.~Metlitski,
  ``$S$-duality of $u(1)$ gauge theory with $\theta =\pi$ on non-orientable manifolds: Applications to topological insulators and superconductors,''
  arXiv:1510.05663 [hep-th].

\bibitem{Seiberg:2016rsg}
  N.~Seiberg and E.~Witten,
  ``Gapped Boundary Phases of Topological Insulators via Weak Coupling,''
  PTEP {\bf 2016}, no. 12, 12C101 (2016)
  doi:10.1093/ptep/ptw083
  [arXiv:1602.04251 [cond-mat.str-el]].

\bibitem{Tachikawa:2016xvs}
  Y.~Tachikawa and K.~Yonekura,
  ``Gauge interactions and topological phases of matter,''
  PTEP {\bf 2016}, no. 9, 093B07 (2016)
  doi:10.1093/ptep/ptw131
  [arXiv:1604.06184 [hep-th]].

\bibitem{Freed:2016rqq}
  D.~S.~Freed and M.~J.~Hopkins,
  ``Reflection positivity and invertible topological phases,''
  arXiv:1604.06527 [hep-th].

\bibitem{Witten:2016cio}
  E.~Witten,
  ``The `Parity' Anomaly On An Unorientable Manifold,''
  Phys.\ Rev.\ B {\bf 94}, no. 19, 195150 (2016)
  doi:10.1103/PhysRevB.94.195150
  [arXiv:1605.02391 [hep-th]].

\bibitem{Yonekura:2016wuc}
  K.~Yonekura,
  ``Dai-Freed theorem and topological phases of matter,''
  JHEP {\bf 1609}, 022 (2016)
  doi:10.1007/JHEP09(2016)022
  [arXiv:1607.01873 [hep-th]].

\bibitem{Hasebe:2016tjg}
  K.~Hasebe,
  ``Higher (Odd) Dimensional Quantum Hall Effect and Extended Dimensional Hierarchy,''
  Nucl.\ Phys.\ B {\bf 920}, 475 (2017)
  doi:10.1016/j.nuclphysb.2017.03.017
  [arXiv:1612.05853 [hep-th]].

\bibitem{Yu:2017uqt}
  Y.~Yu, Y.~S.~Wu and X.~Xie,
  ``Bulk–edge correspondence, spectral flow and Atiyah–Patodi–Singer theorem for the Z2 -invariant in topological insulators,''
  Nucl.\ Phys.\ B {\bf 916}, 550 (2017)
  doi:10.1016/j.nuclphysb.2017.01.018
  [arXiv:1607.02345 [cond-mat.mes-hall]].



\bibitem{Grabowska:2015qpk} 
  D.~M.~Grabowska and D.~B.~Kaplan,
  ``Nonperturbative Regulator for Chiral Gauge Theories?,''
  Phys.\ Rev.\ Lett.\  {\bf 116}, no. 21, 211602 (2016)
  doi:10.1103/PhysRevLett.116.211602
  [arXiv:1511.03649 [hep-lat]].


\bibitem{Fukaya:2016ofi} 
  H.~Fukaya, T.~Onogi, S.~Yamamoto and R.~Yamamura,
  ``Six-dimensional regularization of chiral gauge theories,''
  PTEP {\bf 2017}, no. 3, 033B06 (2017)
  doi:10.1093/ptep/ptx017
  [arXiv:1607.06174 [hep-th]].


\bibitem{Okumura:2016dsr} 
  K.~i.~Okumura and H.~Suzuki,
  ``Fermion number anomaly with the fluffy mirror fermion,''
  PTEP {\bf 2016}, no. 12, 123B07 (2016)
  doi:10.1093/ptep/ptw167
  [arXiv:1608.02217 [hep-lat]].


\bibitem{Hamada:2017tny} 
  Y.~Hamada and H.~Kawai,
  ``Axial U(1) current in Grabowska and Kaplan’s formulation,''
  PTEP {\bf 2017}, no. 6, 063B09 (2017)
  doi:10.1093/ptep/ptx086
  [arXiv:1705.01317 [hep-lat]].
  
\bibitem{Callan:1984sa} 
  C.~G.~Callan, Jr. and J.~A.~Harvey,
  ``Anomalies and Fermion Zero Modes on Strings and Domain Walls,''
  Nucl.\ Phys.\ B {\bf 250}, 427 (1985).
  doi:10.1016/0550-3213(85)90489-4
  
\bibitem{Kaplan:1992bt} 
  D.~B.~Kaplan,
  ``A Method for simulating chiral fermions on the lattice,''
  Phys.\ Lett.\ B {\bf 288}, 342 (1992)
  doi:10.1016/0370-2693(92)91112-M
  [hep-lat/9206013].
\bibitem{Shamir:1993zy} 
  Y.~Shamir,
  ``Chiral fermions from lattice boundaries,''
  Nucl.\ Phys.\ B {\bf 406}, 90 (1993)
  doi:10.1016/0550-3213(93)90162-I
  [hep-lat/9303005].
\bibitem{Furman:1994ky} 
  V.~Furman and Y.~Shamir,
  ``Axial symmetries in lattice QCD with Kaplan fermions,''
  Nucl.\ Phys.\ B {\bf 439}, 54 (1995)
  doi:10.1016/0550-3213(95)00031-M
  [hep-lat/9405004].


  
\bibitem{Witten:1982fp} 
  E.~Witten,
  ``An SU(2) Anomaly,''
  Phys.\ Lett.\  {\bf 117B}, 324 (1982).
  doi:10.1016/0370-2693(82)90728-6

\bibitem{Hortacsu:1980kv} 
  M.~Hortacsu, K.~D.~Rothe and B.~Schroer,
  ``Zero Energy Eigenstates for the Dirac Boundary Problem,''
  Nucl.\ Phys.\ B {\bf 171}, 530 (1980).
  doi:10.1016/0550-3213(80)90384-3

  
\bibitem{Luscher:2006df} 
  M.~Luscher,
  ``The Schrodinger functional in lattice QCD with exact chiral symmetry,''
  JHEP {\bf 0605}, 042 (2006)
  doi:10.1088/1126-6708/2006/05/042
  [hep-lat/0603029].

  
\bibitem{Neuberger:1997fp} 
  H.~Neuberger,
  ``Exactly massless quarks on the lattice,''
  Phys.\ Lett.\ B {\bf 417}, 141 (1998)
  doi:10.1016/S0370-2693(97)01368-3
  [hep-lat/9707022].
  
\bibitem{Neuberger:1998wv} 
  H.~Neuberger,
  ``More about exactly massless quarks on the lattice,''
  Phys.\ Lett.\ B {\bf 427}, 353 (1998)
  doi:10.1016/S0370-2693(98)00355-4
  [hep-lat/9801031].


\bibitem{Trott}
  M.~Trott, S.~Trott and Ch.~Schnittler,
  ``Normalization, Orthogonality, and Completeness for the Finite Step Potential,''
  Phys.\ Stat.\ Sol.\ (b) {\bf 151}, (1989) K123.


\bibitem{Marachevsky:2003zb} 
  V.~N.~Marachevsky and D.~V.~Vassilevich,
  ``Chiral anomaly for local boundary conditions,''
  Nucl.\ Phys.\ B {\bf 677}, 535 (2004)
  doi:10.1016/j.nuclphysb.2003.11.009
  [hep-th/0309019].

\bibitem{Beneventano:2004yy} 
  C.~G.~Beneventano and E.~M.~Santangelo,
  ``Spectral functions of the Dirac operator under local boundary conditions,''
  in {\it Mathematical physics frontiers},
  edited by C. V. Benton (Nova Science Publishers, Hauppauge, New York, 2004)
  [hep-th/0405221].

  
  
\bibitem{Hasenfratz:1998ri} 
  P.~Hasenfratz, V.~Laliena and F.~Niedermayer,
  ``The Index theorem in QCD with a finite cutoff,''
  Phys.\ Lett.\ B {\bf 427}, 125 (1998)
  doi:10.1016/S0370-2693(98)00315-3
  [hep-lat/9801021].

\bibitem{Golterman:1992ub} 
  M.~F.~L.~Golterman, K.~Jansen and D.~B.~Kaplan,
  ``Chern-Simons currents and chiral fermions on the lattice,''
  Phys.\ Lett.\ B {\bf 301}, 219 (1993)
  doi:10.1016/0370-2693(93)90692-B
  [hep-lat/9209003].

\bibitem{AlvarezGaume:1985di} 
  L.~Alvarez-Gaume, S.~Della Pietra and V.~Della Pietra,
  ``The Effective Action for Chiral Fermions,''
  Phys.\ Lett.\ B {\bf 166}, 177 (1986).
  doi:10.1016/0370-2693(86)91373-0

\bibitem{DellaPietra:1986qd} 
  S.~Della Pietra, V.~Della Pietra and L.~Alvarez-Gaume,
  ``The Chiral Determinant and the $\eta$ Invariant,''
  Commun.\ Math.\ Phys.\  {\bf 109}, 691 (1987).
  doi:10.1007/BF01208963

  
\bibitem{Kaplan:1995pe} 
  D.~B.~Kaplan and M.~Schmaltz,
  ``Domain wall fermions and the eta invariant,''
  Phys.\ Lett.\ B {\bf 368}, 44 (1996)
  doi:10.1016/0370-2693(95)01485-3
  [hep-th/9510197].

\bibitem{Hashimoto:2017tuh} 
  K.~Hashimoto, X.~Wu and T.~Kimura,
  ``Edge states at an intersection of edges of a topological material,''
  Phys.\ Rev.\ B {\bf 95}, no. 16, 165443 (2017)
  doi:10.1103/PhysRevB.95.165443
  [arXiv:1702.00624 [cond-mat.mes-hall]].

  
  \bibitem{Apostol}
  T. M. Apostol, ``Introduction to Analytic Number Theory,''
  Undergraduate Texts in Mathematics. Springer-Verlag, New York-Heidelberg, 1976.



  
\end{thebibliography}
\end{document}